\documentclass[11pt,draftclsnofoot,onecolumn]{IEEEtran}
\usepackage[T1]{fontenc}
\usepackage{xspace,amsmath,amssymb,amsfonts,epsfig,syntonly,bbm,dsfont}
\usepackage{cite,bm,color,url,textcomp,empheq,boxedminipage}
\usepackage{algorithmicx,algorithm,algpseudocode}
\usepackage{epstopdf}
\usepackage{empheq}
\usepackage{graphicx,graphics,subfigure}
\usepackage{multirow,multicol}
\usepackage{psfrag}
\usepackage{stfloats}

\newtheorem{theorem}{{{Theorem}}}
\newtheorem{lemma}{{Lemma}}
\newtheorem{proposition}{{Proposition}}
\newtheorem{example}{{Example}}
\newtheorem{remark}{{Remark}}

\DeclareMathOperator*{\argmax}{arg\,max}
\DeclareMathOperator*{\argmin}{arg\,min}

\long\def\symbolfootnote[#1]#2{\begingroup
\def\thefootnote{\fnsymbol{footnote}}
\footnote[#1]{#2}\endgroup}

\psfull

\begin{document}
\title{Optimal Energy Allocation and Task Offloading Policy for Wireless Powered Mobile Edge Computing Systems}

\author{{Feng Wang, Jie Xu, and Shuguang Cui}
\thanks{F. Wang is with the School of Information Engineering, Guangdong University of Technology, Guangzhou 510006, China (e-mail: fengwang13@gdut.edu.cn).}

\thanks{J. Xu is with the Future Network of Intelligence Institute (FNii) and the School of Science and Engineering, The Chinese University of Hong Kong (Shenzhen), Shenzhen 518172, China (email: xujie@cuhk.edu.cn). J. Xu is the corresponding author.}

\thanks{S. Cui is with the Shenzhen Research Institute of Big Data and the Future Network of Intelligence Institute (FNii), The Chinese University of Hong Kong (Shenzhen), Shenzhen 518172, China (e-mail: robert.cui@gmail.com).}

 \vspace{-1.5cm}
}

\maketitle

\begin{abstract}
This paper studies a single-user wireless powered mobile edge computing (MEC) system, in which one multi-antenna energy transmitter (ET) employs energy beamforming for wireless power transfer (WPT) towards the user, and the user relies on the harvested energy to locally execute a portion of tasks and offload the other portion to an access point (AP) integrated with an MEC server for remote execution. Different from prior works considering static wireless channels and computation tasks at the user, this paper considers both energy and task causality constraints due to the channel fluctuations and dynamic task arrivals over time. Towards an energy-efficient joint-WPT-MEC design, we minimize the total transmission energy consumption at the ET over a particular finite horizon while ensuring the user's successful task execution, by jointly optimizing the transmission energy allocation at the ET for WPT and the task allocation at the user for local computing and offloading over a particular finite horizon. First, to characterize the fundamental performance limit, we consider the {\em offline optimization} by assuming that the perfect knowledge of channel state information (CSI) and task state information (TSI) (i.e., task arrival timing and amounts) is known {\em a-priori}. In this case, we obtain the well-structured optimal solution to the energy minimization problem by using convex optimization techniques. The optimal solution shows that in the scenario with static channels, the ET should allocate the transmission energy uniformly over time, and the user should employ staircase task allocation for both local computing and offloading, with the number of executed task input-bits monotonically increasing over time. It also shows that in the scenario with time-varying channels, the ET should transmit energy sporadically at slots with causally dominating channel power gains, and the user should apply the staircase task allocation for local computing and staircase water-filling task allocation for offloading with monotonically increasing computation levels over time. Next, inspired by the structured offline solutions obtained above, we develop heuristic {\em online designs} for the joint energy and task allocation when the knowledge of CSI/TSI is only causally known. Finally, numerical results show that the proposed joint energy and task allocation designs achieve significantly smaller energy consumption than benchmark schemes with only local computing or full offloading at the user, and the proposed heuristic online designs perform close to the optimal offline solutions and considerably outperform the conventional myopic designs.
\end{abstract}


\section{Introduction}
 The integration of mobile edge computing (MEC) \cite{MH18,Chen16,JunZhang17,Chiang16,Mach17,Taleb17,Cao18,Feng19,Yuan19} and wireless power transfer (WPT) \cite{Zeng17,Clerckx,Rui-SWIPT,Bi15} has recently emerged as a viable and promising solution to empower a large number of low-power internet-of-things (IoT) wireless devices (such as sensors, meters, cameras, and wearables), with enhanced and sustainable communication and computation performance. In such wireless powered MEC systems, energy transmitters (ETs) and MEC servers are deployed at the mobile network edge, either separated or co-located with access points (APs) or base stations (BSs) therein. Accordingly, the ETs are allowed to wirelessly supply the on-demand energy for wireless devices' central processing units (CPUs) and radio transceivers via WPT in a fully controllable manner. Relying on the harvested energy, wireless devices offload (partial or full) computation tasks to APs within their radio access for MEC servers' computing therein and execute the rest tasks locally. By exploiting both benefits of MEC and WPT, the wireless powered MEC is able to significantly prolong the wireless network lifetime and even achieve sustainable (battery-free) network operation, with enhanced computation and communication capability at end wireless devices. Therefore, this wireless powered MEC technique is envisioned to enable low-power wireless devices to perform the interactive communication and computation in future (battery-free) IoT applications \cite{Mach17,Taleb17,Chiang16}. For example, in an environment sensing application, the wireless devices rely on the harvested RF energy to collect the raw data (e.g., temperature, humidity, pressure, intensity, tactility), pre-process such data, and then send them to the AP for post-processing.

 The wireless powered MEC systems face various new technical challenges due to the coupling of the wireless energy supply and the communication and computation demand at users. This thus calls for a new design framework to jointly optimize the WPT at ETs and the task execution via local computing and offloading at users, for maximizing the system performance. In the literature, the authors in \cite{You16} first considered a single-user wireless powered MEC system with co-located ET and MEC server, with the objective of maximizing the probability of successfully computing given tasks at the user. Furthermore, the authors in \cite{Feng18} studied multiuser wireless powered MEC systems under a time-division multiple access (TDMA) protocol for multiuser computation offloading, in which the overall energy consumption (including the transmission energy for WPT at the ET and the remote computing energy at the MEC server) is minimized subject to both energy neutrality and task completion constraints at these users. Moreover, \cite{Bi18} studied the computation rate maximization in multiuser wireless powered MEC systems with binary offloading. The authors in \cite{Hu18} further considered a wireless powered relaying system for MEC, where one relay node utilizes the harvested wireless energy to help the source node's task offloading via relaying. In addition, \cite{Wu18} studied the computation rate maximization in wireless powered user cooperative computation systems, in which the source user node utilizes its harvested energy from the ET to offload its tasks to multiple peer user nodes via device-to-device (D2D) links and each peer node also opportunistically harvests the harvested wireless energy for cooperative computation.

 Despite such research progress, these prior works \cite{You16,Feng18,Bi18,Hu18,Wu18} focused on one-shot optimization under static wireless channels and given computation tasks at users, in which the time-dynamics in both WPT and task arrivals are overlooked. In practical wireless powered MEC systems, nonetheless, both wireless energy and computation task arrivals at users may fluctuate significantly over time, due to the randomness in wireless channels and the bursty nature of computation traffics, respectively. Therefore, both energy and task causality constraints are imposed at users, i.e., the energy (or task) amount cumulatively consumed (or executed) at any time instant cannot exceed that cumulatively harvested (or arrived) at that time. Under these new constraints, how to adaptively manage the ET's wireless energy supply over time-varying channels to support users' dynamic computation demands with random task arrivals is a fundamental but challenging problem that remains not well addressed yet. This thus motivates the current work.

 In this paper, as a starting point to gain essential insights for wireless powered MEC designs with random task arrivals, we consider a basic single-user system setup that consists of a multi-antenna ET, a single-antenna AP integrated with an MEC server, and a single-antenna user node with dynamic task arrivals over time. The ET employs the energy beamforming to wirelessly charge the user, and the user relies on the harvested energy to execute its computation tasks via locally computing a portion of them and offloading the other portion to the AP. To avoid the co-channel interference, we assume that the WPT from the ET to the user and the task offloading from the user to the AP are implemented simultaneously over orthogonal frequency bands. We focus on a particular finite time horizon consisting of multiple slots. Suppose that the user's computation tasks arrive at the beginning of each slot and all tasks need to be successfully executed before the end of this horizon. Towards an energy efficient joint-WPT-MEC design, our objective is to minimize the transmission energy consumption for WPT at the ET subject to the energy and task causality constraints at the user, by jointly optimizing the energy allocation for WPT at the ET and task allocation for local computing and offloading at the user over time. The main results of this paper are summarized as follows.

\begin{itemize}
\item First, in order to characterize the fundamental performance limit, we consider the offline optimization by assuming that the perfect knowledge of channel state information (CSI) and task state information (TSI) (i.e., task arrival timing and amounts) is known {\em a-priori}. In this case, the energy minimization problem corresponds to a convex optimization problem. Then, we handle this problem by first considering the special scenario with static channels. In this scenario, we obtain a well-structured optimal solution by leveraging the Karush-Kuhn-Tucker (KKT) optimality conditions. The optimal solution shows that the ET should allocate the transmission energy uniformly over time, and the user should employ staircase task allocation for both local computing and offloading, with the number of executed task input-bits monotonically increasing over time.
\item Next, we consider the general scenario with time-varying channels, in which the energy minimization problem becomes more challenging to solve. In this scenario, we show that this problem can be decomposed into two subproblems for the ET's energy allocation and the user's task allocation, respectively. Accordingly, we obtain the well-structured optimal solution via convex optimization techniques. It is shown that the ET should transmit energy sporadically at slots with causally dominating channel power gains, and the user should apply the staircase task allocation for local computing and the staircase water-filling for offloading with monotonically increasing computation levels over time.
\item In addition, we also consider the online optimization when the knowledge of CSI and TSI is causally known, i.e., at each time slot, only the past and present CSI/TSI is available but the future CSI/TSI is unknown. Inspired by the structured optimal offline solutions obtained above, we develop heuristic online designs for the joint energy allocation (for WPT) at the ET and task allocation (for local computing and offloading) at the user, under both scenarios with static and time-varying channels.
\item Finally, we provide numerical results to validate the performance of our proposed designs. It is shown that in both static and time-varying channel scenarios, the optimal offline solutions achieve significantly smaller energy consumption than benchmark schemes with only local computing or full offloading at the user, while the proposed heuristic online designs perform close to the offline solutions and considerably outperform the conventional myopic designs.
\end{itemize}

It is worth emphasizing that the proposed joint energy and task allocation designs in wireless powered MEC systems are different from the task allocation in energy harvesting powered MEC systems \cite{Mao16JASC,Mao17TWC,Jie17Cog,Liang19TVT,Liang19IoT}, the energy allocation in energy harvesting\cite{Rui12} or wireless powered communication systems \cite{XunZhou16}, and the power usage for bursty data packet transmissions \cite{Xin13}. First, unlike \cite{Mao16JASC,Mao17TWC,Jie17Cog} considering random and uncontrollable energy arrivals from ambient renewable sources (e.g., solar and wind energy), this paper considers the fully controllable energy supply from WPT at the ET, in which the energy allocation for WPT is an additional design degree of freedom for optimizing the system computation performance. Next, in contrast to \cite{Rui12} and \cite{XunZhou16} with only communication energy consumption considered, this paper focuses on both communication (for offloading) and (local) computation energy consumptions at the user, thus making the demand side management (with task allocation) more challenging. Specifically, the work in \cite{XunZhou16} considered a wireless powered orthogonal frequency division multiplexing (OFDM) communication system with multiple users, in which the subcarrier and power allocations over time for WPT and wireless information transmission are jointly optimized to maximize the uplink sum rate at users. By contrast, this paper considers an energy-efficient single-user wireless powered MEC system (instead of communication only) under randomly arrived tasks at the user, in which we aim to minimize the ET's energy consumption under a given computation latency constraint. By optimally solving the energy minimization problem offline, we reveal the optimal energy and task allocation policies over time in well structures, and further develop heuristic online designs inspired by the obtained offline solution. In addition, compared to \cite{Xin13} which revealed a multi-level water-filling form for the optimal power usage in energy-efficient bursty data packet transmission, this paper further considers the energy causality due to the user's computation is powered by the controllable WPT over wireless channels. Furthermore, it is also worth noticing that our prior work\cite{ICC19} addressed the system energy minimization problem in multiuser wireless powered MEC systems with co-located ET and MEC server at the AP subject to energy and task causality constraints at each user, in which joint energy and task allocation is optimized offline via standard convex optimization techniques. By contrast, in this paper we consider a different setup with the ET and MEC server separately located, under which the optimal offline solutions are obtained in well-structured forms to gain more design insights (instead of only numerical algorithms in \cite{ICC19}) in both static and time-varying channel scenarios, and new heuristic online designs are also proposed to facilitate practical implementation.

The remainder of the paper is organized as follows. Section II introduces the single-user wireless powered MEC system model and formulates the joint energy and task allocation problem of interest. Sections III and IV present the optimal offline solutions to the joint energy and task allocation problem in the scenarios with static and time-varying channels, respectively. Building upon the optimal offline designs, Section V presents heuristic online designs for the joint energy and task allocation. Section VI provides numerical results to demonstrate the effectiveness of the proposed designs, followed by the concluding remark in Section VII.

{\em Notation:} For an arbitrary-size matrix $\bm M$, $\bm M^H$ denotes the conjugate transpose. $\mathbb{C}^{x\times y}$ denotes the space of $x\times y$ matrices with complex entries. $\|\bm z\|$ denotes the Euclidean norm of a complex vector $\bm z$, $|z|$ denotes the absolute value of a complex scalar $z$, and $|{\cal X}|$ denotes the cardinality of a set ${\cal X}$. $\bm{I}$ and $\bm 0$ denote an identity matrix and an all-zeros vector/matrix, respectively, with appropriate dimensions; $x\sim {\cal CN}(\mu,\sigma^2)$ denotes the distribution of a circular symmetric complex Gaussian (CSCG) random variable $x$ with mean $\mu$ and variance $\sigma^2$, $x\sim {\cal U}[a,b]$ denotes the distribution of a uniform random variable $x$ within an interval $[a,b]$, and $\sim$ stands for ``distributed as''; $\mathbb{E}[\cdot]$ denotes the statistical expectation. Furthermore, we define $[x]^+ \triangleq \max(x,0)$. The main symbols and notations are summarized in Table I.

\begin{table}
\centering
\caption{Summary of the main symbols and notations}
\begin{tabular}{|l|l|l|l|}
\hline
$M$ & Number of the ET's antennas. & $A_i$ & Number of task input-bits arrived at slot $i$ \\
$N$, $i$  & Number and index of time slots. & $B$ & System bandwidth for offloading \\
 $\hat{\bm h}_i$, $h_i$ & Channel vector and power gain for WPT at slot $i$ & $T$ & Duration of the time horizon \\
 $\hat{g}_i$, $g_i$ & Channel coefficient and power gain for offloading at slot $i$ & $\tau$ & Duration of one slot \\
$C$ & Number of CPU cycles to execute one task input-bit & $h_i^\prime$ & Effective WPT channel power gain at slot $i$ \\
${\cal N}_{\rm CDS}$ & Set of causality dominating slots & $\sigma^2$ & Power of the AWGN at AP \\
${\cal N}_{\rm TS}$ & Set of transition slots & $\eta$ & RF-to-DC conversion efficient for WPT \\
 $p_i$ & Power of ET's transmission for WPT at slot $i$ & $\zeta$ & CPU's switched capacitance coefficient \\
$\ell_i$ & Number of task input-bits for local computing at slot $i$ & $d_i$ & Number of task input-bits for offloading at slot $i$ \\
$\omega_i$ & Computation level at slot $i$ under time-varying channels & $\nu_i$ & Computation level at slot $i$ under static channels  \\
\hline
\end{tabular}
\end{table}

\section{System Model and Problem Formulation}\label{Sec:System}

\begin{figure}
  \centering
  \includegraphics[width = 4.0in]{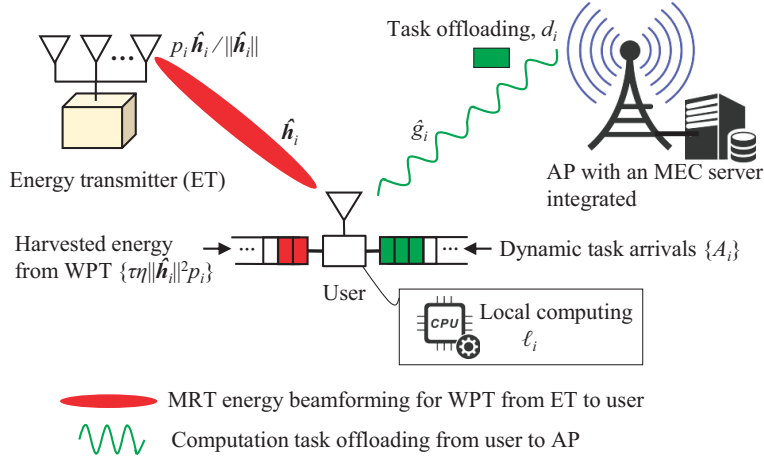}
 \caption{An illustration of the single-user wireless powered MEC system.} \label{fig.system-model}
\end{figure}

 As shown in Fig.~\ref{fig.system-model}, this paper considers a basic single-user wireless powered MEC system, which consists of an ET equipped with $M>1$ antennas, a single-antenna AP integrated with an MEC server\footnote{Note that the MEC server represents an integral part of the virtualized computation resources and hosts the MEC applications running at the virtual machines (VMs) on top of the virtualization infrastructure\cite{Mach17,Taleb17}.}, and a single-antenna user with dynamically arrived tasks to be executed. In this system, the ET employs the energy beamforming to charge the user over the air, and the user relies on the harvested energy for task execution via computing locally or offloading to the AP. Suppose that the WPT from the ET to the user and the computation offloading from the user to the AP are implemented simultaneously over orthogonal frequency bands. In particular, we focus on a finite time horizon with duration $T>0$, which is divided into $N$ time slots each with identical duration $\tau=T/N$. Let ${\cal N}\triangleq \{1,\ldots,N\}$ denote the set of the $N$ slots. At the beginning of each slot $i\in{\cal N}$, let $A_{i}\geq 0$ denote the number of task input-bits arrived at the user. In this paper, we assume that the computation tasks are subject to a common computation deadline at the end of this horizon, i.e., the user needs to successfully execute these tasks before the horizon end. This assumption is motivated by some practical IoT applications. For example, the user (e.g., a sensor node) may need to take actions after performing several sensing tasks over time. In this case, these different tasks should have a common computation deadline.

\subsection{Task Execution at User}
In this subsection, we consider the task execution at the user via local computing and task offloading, respectively. At each slot $i\in{\cal N}$, let $\ell_i\geq 0$ and $d_i\geq 0$ denote the number of task input-bits which are computed locally and offloaded to the AP, respectively.\footnote{As an initial investigation for unified WPT-MEC designs with dynamic task arrivals, we assume all the arrived tasks under consideration are partitionable in this paper, such that the user can arbitrarily partition its computation tasks into two parts for local computing and offloading to the AP, respectively. This generally serves as a performance upper bound for other scenarios when the computation tasks are not partitionable\cite{Mach17,Taleb17,Chiang16,Bi18,Feng19} or when the computation tasks can only be partitioned into dependent sub-tasks \cite{JunZhang17}.} The task execution at the user is subject to the task causality constraints, i.e., until each slot $i\in{\cal N}$, the number of task input-bits cumulatively executed via both local computing and task offloading (i.e., $\sum_{j=1}^i (\ell_j+d_j)$) {\em cannot} exceed that cumulatively arrived (i.e., $\sum_{j=1}^i A_j$). Therefore, we have
\begin{align}\label{eq.Si-user}
\sum_{j=1}^i (\ell_{j}+ d_{j})\leq  \sum_{j=1}^i A_{j} ,~~ \forall i\in{\cal N}.
\end{align}
In addition, since the user needs to successfully accomplish the task execution before the end of the last slot $N$, we have the task completion constraint as
\begin{align} \label{eq.task-comp}
\sum_{j=1}^N (\ell_{j} + d_{j}) = \sum_{j=1}^N A_{j}.
\end{align}

First, we consider the user's local computing for executing the $\ell_{i}$ task input-bits at each slot $i\in{\cal N}$, where the local computing time per slot is equal to the slot length $\tau$. Let $C\geq 0$ denote the number of central processing unit (CPU) cycles required for executing one task input-bit at the user, which generally depends on the types of applications and the user's CPU architecture\cite{Burd96}. Accordingly, a total of $C\ell_{i}$ CPU cycles are required for the user's local computing. By applying the dynamic voltage and frequency scaling (DVFS) technique, in order to maximize the energy efficiency for local computing, the user should adopt a constant CPU frequency $C\ell_{i}/\tau$ at each slot $i\in{\cal N}$\cite{Feng18}. In this case, the user's energy consumption for local computing at slot $i\in{\cal N}$ is expressed as \cite{Burd96}
\begin{align}\label{eq.Eki}
E^{\rm loc}(\ell_i) = C\ell_{i} \zeta \left(\frac{C\ell_{i}}{\tau} \right)^2 =\frac{\zeta C^3 \ell_{i}^3}{\tau^2},
\end{align}
where $\zeta>0$ denotes the effective switched capacitance coefficient depending on the user's CPU chip architecture.

Next, we consider the user's computation offloading of the $d_i$ task input-bits at each slot $i\in{\cal N}$. Note that the user's computation offloading procedure consists of the three consecutive phases: task offloading, remote execution, and result downloading. As commonly considered in the MEC literature (see, e.g., \cite{Feng18,You16,Bi18}), in this paper we focus on the first phase of task offloading, by modelling both the task remote execution time and results downloading time as two constants that are considerably smaller than the task offloading time. This is practically reasonable for many computation-intensive IoT applications (e.g., image/video/voice recognition, file scanning, data analysis, multisensory information processing, etc.), since the AP (integrated with the MEC server) normally has sufficiently rich computation and computation resources than the user, and the size of the computation results is generally much smaller than that of the task input-bits. For the purpose of idea exposition, the user's task offloading time at each slot is set to be the slot length in this paper.

Let $q_i$, $g_i$, and $B$ denote the user's transmission power, the channel power gain, and the system bandwidth for task offloading from the user to the AP, respectively. The transmission rate for offloading (in bits-per-second) from the user to the AP at slot $i\in{\cal N}$ is expressed as
\begin{align}
r_i = B\log_2\left(1+\frac{g_iq_i}{\Gamma\sigma^2}\right),
\end{align}
where $\Gamma\geq 1$ denotes the signal-to-noise ratio (SNR) gap due to the practical adaptive modulation and coding (AMC) scheme employed at the user\cite{Goldsmith} and $\sigma^2$ denotes the power of the additive white Gaussian noise (AWGN) at the AP receiver. For notational convenience, we define $\Gamma\triangleq 1$ in the sequel. In this case, we have $d_i= r_i\tau $ in order for the user to offload the $d_i$ task input-bits to the AP. As a result, the user's transmission energy consumption for task offloading at each slot $i\in{\cal N}$ is given by
\begin{align}\label{eq.Eoff}
E_{i}^{\rm offl}(d_i) =  \tau q_i =\frac{\tau\sigma^2}{g_i}\big(2^{\frac{d_{i}}{\tau B}}-1\big).
\end{align}
Notice that both $E^{\rm loc}(\ell_i)$ in \eqref{eq.Eki} and $E_i^{\rm offl}(d_i)$ in \eqref{eq.Eoff} are convex functions with respect to $\ell_i \geq 0$ and $d_i\geq 0$, respectively.

\subsection{Energy Beamforming for WPT at ET}
 In this subsection, we consider the energy beamforming for WPT at the ET to wirelessly charge the user node. At each slot $i\in{\cal N}$, let $s_i$ denote the energy-bearing signal at the ET, where $\mathbb{E}[|s_i|^2]=1$ is assumed without loss of generality. Also, let ${\bm w}_i\in\mathbb{C}^{M\times1}$ (with $\|{\bm w}_i\|=1$) and $p_i\geq 0$ denote the energy beamforming vector and transmission power at the ET, respectively. Then, the transmitted energy signal of the ET is ${\bm x}_i=\sqrt{p_i}{\bm w}_is_i$. Let $\hat{\bm h}_i\in\mathbb{C}^{M\times1}$ denote the channel vector from the ET to the user for downlink WPT. The harvested energy by the user in this slot is then given by $\tau\eta p_i\|{\bm w}^H_i\hat{\bm h}_i\|^2$, where $0<\eta\leq 1$ denotes the RF-to-DC (direct current) energy conversion efficiency.\footnote{The RF-to-DC energy conversion is generally a nonlinear process, and the energy conversion efficiency highly depends on both the input-RF power and signal waveform\cite{Zeng17,Clerckx,Zhou18WPT,Zhou18CL,Zhou19NOMA}. To our best knowledge, there still lacks a generic RF-to-DC energy conversion model in the literature. Nevertheless, it is shown in \cite{Rui-SWIPT,Zeng17,Clerckx} that when the input RF power is smaller than a certain saturation level, the RF-to-DC energy conversion can be nicely approximated as a linear process. For the purpose of initial investigation, we assume that the RF-to-DC energy conversion at the user works at the linear regime in this paper. In  practice, this can be implemented by the ET properly adjusting the transmit power under our single-user scenario of interest.}

Motivated by the above discussions, we herein adopt a linear energy harvesting model in this paper, by assuming that the ET can properly adjust its transmission power level for WPT such that the user's received RF power is always within the linear regime for RF-to-DC conversion at the rectifier\cite{Zeng17,Clerckx}. Furthermore, we assume that the ET employs the maximum ratio transmission (MRT) energy beamforming to maximize the transferred energy towards the user\cite{Rui-SWIPT} by setting ${\bm w}_i= \hat{\bm h}_i/\|\hat{\bm h}_i\|$, $\forall i\in{\cal N}$. As a result, the energy harvested by the user at slot $i\in{\cal N}$ is given by $E_i^{\rm EH}(p_i) = \tau \eta h_i p_i$, where $h_i\triangleq \|\hat{\bm h}_i\|^2$ denotes the channel power gain for WPT from the ET to the user at slot $i$.

Note that the user's local computing and task offloading are both powered by the wireless energy transferred from the ET, thereby achieving sustainable computation and communication. In practice, the harvested energy at each slot $i\in{\cal N}$ can only be utilized at the present and subsequent time slots. In this case, the user is subject to the so-called {\em energy causality} constraints\cite{Rui12,XunZhou16}, i.e., at each slot $i\in{\cal N}$, the cumulatively consumed energy amount (for local computing and task offloading) at the user (i.e., $\sum_{j=1}^i ( E^{\rm loc}(\ell_j)+E_j^{\rm offl}(d_j))$) cannot exceed that cumulatively harvested from the ET at that slot (i.e., $\sum_{j=1}^{i} E_j^{\rm EH}(p_j)$). As a result, we have
\begin{align}\label{eq.EHi}
\sum_{j=1}^i ( E^{\rm loc}(\ell_j)+E_j^{\rm offl}(d_j) ) \leq \sum_{j=1}^{i} E_j^{\rm EH}( p_j), ~~\forall i\in {\cal N}.
\end{align}

\subsection{Problem Formulation}
In this paper, we are interested in minimizing the transmission energy consumption for WPT at the ET while achieving sustainable operation for the user's communication and computation\footnote{In this work, the energy consumption at the AP/MEC server incurred by offloading is not considered, since this term can be generally modeled as a constant. In practice, after the user offloads tasks to the AP, the MEC server therein normally uses the peak (and thus fixed) CPU frequency to execute these offloaded tasks with minimum delay\cite{Mach17,Taleb17,JunZhang17}. In this case, the potential computation energy consumption reduction by controlling the AP's remote execution for MEC is generally a negligible part as compared to the AP's operational energy consumption for e.g., circuit, signal processing, and cooling.}. In particular, our objective is to minimize the ET's transmission energy consumption (i.e., $\sum_{i=1}^{N}\tau p_i$), subject to the user's task causality constraints in \eqref{eq.Si-user}, task completion constraint in \eqref{eq.task-comp}, and energy causality constraints in \eqref{eq.EHi}. The design variables include the energy allocation $\{p_i\}$ at the ET, as well as the task allocation of $\{\ell_i\}$ for local computing and $\{d_i\}$ for offloading at the user. Mathematically, the energy minimization problem of interest is formulated as
\begin{subequations}\label{eq.prob0}
\begin{align}
{\rm ({\cal P}1)}: & \min_{\{p_i\geq 0,\ell_{i}\geq 0,d_{i}\geq 0\}}  ~ \sum_{i=1}^{N} \tau p_i \\
  &\quad\quad~~ {\rm s.t.}~~ \sum_{j=1}^i\left(E^{\rm loc}(\ell_j)+E_j^{\rm offl}(d_j)\right) \leq \sum_{j=1}^i \tau \eta h_j p_j,~~\forall i\in{\cal N}\label{eq.energy-causality}\\
&\quad\quad\quad\quad~~  \sum_{j=1}^i (\ell_j + d_j) \leq \sum_{j=1}^i A_j,~~   \forall i\in{\cal N}\setminus\{N\} \label{eq.task-causality}\\
&\quad\quad\quad\quad~~  \sum_{j=1}^N (\ell_j + d_j) = \sum_{j=1}^N A_j.\label{eq.task-completion}
\end{align}
\end{subequations}
Notice that the solution of problem (${\cal P}1$) critically depends on the availability of the knowledge of CSI (i.e., $\{h_i\}$ and $\{g_i\}$) and TSI (i.e., $\{A_i\}$). In this paper, we first focus on the offline optimization with non-causal CSI and TSI, i.e., the CSI of $\{h_i\}_{i=1}^N$ and $\{g_i\}_{i=1}^N$ and the TSI of $\{A_i\}_{i=1}^N$ are perfectly known {\em a-priori}. The offline optimization serves as the fundamental performance upper bound (i.e., the ET's transmission energy consumption lower bound) for all the designs under imperfect and/or causally known CSI/TSI, which thus helps draw essential insights to motivate practical designs. In this case, since the computation energy consumption functions $E^{\rm loc}(\ell_i)$ and $E_i^{\rm offl}(d_i)$ are convex functions with respect to $\ell_i\geq 0$ and $d_i\geq 0$, respectively, problem (${\cal P}1$) is a convex optimization problem that can be efficiently solved by standard convex optimization techniques\cite{Boyd2004}. In Sections III and IV, we will obtain well-structured optimal solutions to problems (${\cal P}$1) in the scenarios with static and time-varying channels, respectively. Next, inspired by the optimal offline solutions, in Section V we will consider the online optimization of problem (${\cal P}1$) with causal CSI/TSI available, i.e., at each slot $i\in{\cal N}$, only the CSI of $\{h_j\}_{j=1}^i$ and $\{g_j\}_{j=1}^i$ and the TSI of $\{A_j\}_{j=1}^i$ for the previous and present slots are perfectly known, but $\{h_j\}_{j=i+1}^N$, $\{g_j\}_{j=i+1}^N$, and $\{A_j\}_{j=i+1}^N$ for future slots are unknown.

\section{Optimal Energy and Task Allocation Under Static Channels}
In this section, we consider the offline optimization of problem (${\cal P}$1) under the special scenario with static channels, where $h_i= h$ and $g_i=g$, $\forall i\in{\cal N}$. In this scenario, we define $E^{\rm offl}(x) \triangleq \frac{\tau \sigma^2}{g}(2^{\frac{x}{\tau B}}-1)$ for notational convenience. Accordingly, the energy minimization problem (${\cal P}1$) is reduced as
\begin{subequations}\label{eq.prob2}
\begin{align}
 {\rm ({\cal P}2)}: ~& \min_{\{p_i\geq 0,\ell_{i}\geq 0,d_{i}\geq 0\}}  \sum_{i=1}^{N} \tau p_i \\
 &\quad\quad~ {\rm s.t.}~~  \sum_{j=1}^i \left( E^{\rm loc}(\ell_j) + E^{\rm offl}(d_j) \right) \leq \sum_{j=1}^i \tau \eta h p_j ,~~ \forall i\in{\cal N}\\
& \quad\quad\quad\quad~ \eqref{eq.task-causality}~{\rm and}~\eqref{eq.task-completion}.\notag
\end{align}
\end{subequations}
Let $\{p_i^{**},\ell_i^{**},d_i^{**}\}$ denote the optimal solution to problem (${\cal P}$2). In the following, we obtain the well-structured optimal solution $\{p_i^{**},\ell_i^{**},d_i^{**}\}$ to problem (${\cal P}2$). To start with, we establish the following lemma on the tightness of the $N$-th constraint in (\ref{eq.prob2}b) at the optimality.

 \begin{lemma}\label{lem_tightness}
 At the optimality of problem (${\cal P}2$), the $N$-th constraint in (\ref{eq.prob2}b) must be tight. Equivalently, it must hold that
 \begin{align} \label{eq.N}
 \sum_{i=1}^N \frac{1}{\eta h}\left( E^{\rm loc}(\ell^{**}_i) + E^{\rm offl}(d^{**}_i) \right) = \sum_{i=1}^N \tau p^{**}_i.
 \end{align}
 \end{lemma}
 \begin{IEEEproof}
 We prove Lemma~\ref{lem_tightness} by contradiction. Based on the constraints (\ref{eq.prob2}b) in problem (${\cal P}2$), it must hold that $\sum_{i=1}^N \frac{1}{\eta h}\left( E^{\rm loc}(\ell^{**}_i) + E^{\rm offl}(d^{**}_i) \right) \leq \sum_{i=1}^N \tau p^{**}_i$. Therefore, we suppose that the optimal solution to problem (${\cal P}2$) is obtained as $\{p_i^\prime,\ell_i^{**},d_i^{**}\}$, which satisfies that $\sum_{j=1}^N \left( E^{\rm loc}(\ell^{**}_j) + E^{\rm offl}(d^{**}_j) \right) < \sum_{j=1}^N \tau \eta h p^{\prime}_j$ and $\sum_{j=1}^i \left( E^{\rm loc}(\ell^{**}_j) + E^{\rm offl}(d^{**}_j) \right) < \sum_{j=1}^i \tau \eta h p^{\prime}_j$ for any $i\in{\cal N}\setminus\{N\}$. By setting $p_j^{\prime\prime}=p_j^{\prime}-\Delta$ with $\Delta>0$ for a certain $j\in{\cal N}$ and $p_i^{\prime\prime}=p_i^{\prime}$ for $i\in{\cal N}\setminus\{j\}$, we can always construct a solution $\{p_i^{\prime\prime},\ell_i^{**},d_i^{**}\}$ to problem (${\cal P}2$), which is feasible for constraints (\ref{eq.prob2}b). Furthermore, it is evident that $\sum_{i=1}^N \tau p_i^{\prime\prime} <\sum_{i=1}^N \tau p_i^\prime$. Accordingly, this shows that $\{p_i^\prime,\ell_i^{**},d_i^{**}\}$ is not the optimal solution to problem (${\cal P}2$). Therefore, the optimal solution to problem (${\cal P}2$) must satisfy the $N$-th constraint in~(\ref{eq.prob2}b) with equality, which thus proves Lemma~\ref{lem_tightness}.
 \end{IEEEproof}

 Based on Lemma~\ref{lem_tightness}, the objective function of problem (${\cal P}2$) can be equivalently substituted as the quantity $\sum_{i=1}^N \frac{1}{\eta h}\left( E^{\rm loc}(\ell_i) + E^{\rm offl}(d_i) \right)$. Therefore, the optimal task allocation solution of $\{\ell^{**}_i\}$ and $\{d^{**}_i\}$ to problem (${\cal P}2$) can be obtained by equivalently solving the following total computation energy consumption minimization problem:
\begin{align*}
{({\cal P}2.1)}:~&\min_{\{\ell_{i}\geq 0,d_{i}\geq 0\}}  \sum_{i=1}^N \frac{1}{\eta h}\left(E^{\rm loc}(\ell_i)+E^{\rm offl}(d_i)\right) \notag \\
&\quad~~ {\rm s.t.}~~ \eqref{eq.task-causality}~{\rm and}~\eqref{eq.task-completion}. \notag
\end{align*}
Furthermore, under $\{\ell_i^{**}\}$ and $\{d_i^{**}\}$ obtained in problem (${\cal P}$2.1), it is verified that any energy allocation $\{p_i\}$ satisfying the constraints in (\ref{eq.prob2}b) and the tightness condition in Lemma \ref{lem_tightness} is actually the optimal solution of $\{p_i^{**}\}$ to problem (${\cal P}2$). Therefore, in the following we first obtain $\{\ell^{**}_i\}$ and $\{d_i^{**}\}$ by solving problem (${\cal P}2.1$) and then find $\{p_i^{**}\}$ based on (\ref{eq.prob2}b) and \eqref{eq.N}.

\vspace{-0.2cm}
\subsection{Obtaining Optimal $\{\ell_i^{**}\}$ and $\{d_i^{**}\}$ by Solving Problem (${\cal P}2.1$)}
Note that problem (${\cal P}2.1$) is a convex optimization problem with differential objective and constraint functions. Therefore, strong duality holds between problem (${\cal P}2.1$) and its Lagrange dual problem \cite{Boyd2004}. Let $\mu_i\geq 0$, $\forall i\in{\cal N}\setminus\{N\}$, and $\mu_N\in\mathbb{R}$, denote the Lagrange multipliers associated with the constraints in \eqref{eq.task-causality} and \eqref{eq.task-completion} in problem (${\cal P}2.1$), and $\bar{\theta}_j\geq 0$ and $\underline{\theta}_j\geq 0$, $\forall j\in{\cal N}$, denote the Lagrange multipliers associated with $\ell_j\geq 0$ and $d_j\geq 0$, respectively. The Lagrangian associated with problem (${\cal P}2.1$) is then written as
\begin{align}
{\cal L} = \sum_{i=1}^N \frac{1}{\eta h}\left(E^{\rm loc}(\ell_i)+E^{\rm offl}(d_i)\right) + \sum_{i=1}^{N} \mu_i \Big( \sum_{j=1}^i (\ell_j + d_j) - \sum_{j=1}^i A_j \Big) - \sum_{i=1}^N \bar{\theta}_i\ell_j -\sum_{i=1}^N \underline{\theta}_id_i.
\end{align}
Denote $\{\mu_i^{**},\bar{\theta}_i^{**},\underline{\theta}_i^{**}\}$ as the optimal solution to the Lagrange dual problem of problem (${\cal P}2.1$). Since the convex optimization problem (${\cal P}2.1$) satisfies Slater's condition, the KKT conditions are sufficient and necessary for $\{\ell_i^{**},d^{**}_i\}$ and $\{\mu_i^{**},\bar{\theta}_i^{**},\underline{\theta}_i^{**}\}$ to be the primal and dual optimal solutions to problem (${\cal P}2.1$) with zero duality gap\cite{Boyd2004}. Specifically, it follows that
\begin{subequations}\label{eq.kkt-nonfading}
\begin{align}
& \ell^{**}_i\geq 0, ~d_i^{**}\geq 0, ~\bar{\theta}_i^{**} \geq 0,~ \underline{\theta}_i^{**}\geq 0,~\forall i\in{\cal N},~\mu_j^{**} \geq 0,~\forall j\in{\cal N}\setminus\{N\} \\
&\sum_{j=1}^i(\ell_j^{**}+d_j^{**})- \sum_{j=1}^iA_j\leq 0,~\forall i\in{\cal N}\setminus\{N\},~\sum_{j=1}^N (\ell_j^{**}+d_j^{**}) - \sum_{j=1}^N A_j=0\\
& \bar{\theta}_i^{**} \ell_i^{**} =0,~ \underline{\theta}_i^{**} d_i^{**} =0, ~\mu_i^{**} \Big[\sum_{j=1}^{i}(\ell_j^{**}+d^{**}_j)-\sum_{j=1}^i A_j \Big]=0,~~\forall i\in{\cal N}\\
& \frac{\partial {\cal L}}{\partial \ell_i}\Big|_{\ell_i=\ell_i^{**}} =
\frac{3\zeta C^3(\ell_i^{**})^2}{\eta h\tau^2} - \nu_i - \bar{\theta}_i^{**}= 0,~~\forall i\in{\cal N}\\
& \frac{\partial {\cal L}}{\partial d_i}\Big|_{d_i=d_i^{**}} =  \frac{\sigma^2 \ln 2}{B\eta h g}2^{\frac{d_i^{**}}{\tau B}} - \nu_i - \underline{\theta}_i^{**} = 0,~~\forall i\in{\cal N},
\end{align}
\end{subequations}
where $\nu_i \triangleq -\sum_{j=i}^N\mu^{**}_j$, $\forall i\in{\cal N}$, (\ref{eq.kkt-nonfading}a-b) denote the primal and dual feasible conditions, (\ref{eq.kkt-nonfading}c) denotes the complementary slackness conditions, and (\ref{eq.kkt-nonfading}d) and (\ref{eq.kkt-nonfading}e) state that the gradients of the Lagrangian ${\cal L}$ with respect to $\ell_i$ and $d_i$ vanish at $\ell_i=\ell_i^{**}$ and $d_i=d_i^{**}$, $\forall i\in{\cal N}$, respectively. Based on the KKT conditions in \eqref{eq.kkt-nonfading} together with some algebraic manipulations, one can obtain the optimal solution $\{\ell_i^{**},d_i^{**}\}$ in a closed form to problem (${\cal P}2.1$) in the following theorem.

\begin{theorem}\label{Theo1}
For problem (${\cal P}2.1$), the optimal number of task input-bits $\ell_i^{**}$ for local computing and $d_i^{**}$ for offloading are respectively expressed as
\begin{subequations}\label{eq.lem.p1-1}
\begin{align}
&\ell_i^{**} = \tau\sqrt{\frac{\eta h[\nu_i]^+}{3\zeta C^3}}, ~~\forall i\in{\cal N},\\
&d_i^{**} = \tau B\log_2\Big( \max\Big[ \nu_i\Big/\Big(\frac{\sigma^2 \ln2}{B \eta h g}\Big),~1\Big] \Big),~~\forall i\in{\cal N}.
\end{align}
\end{subequations}
\end{theorem}

\begin{IEEEproof}
First, we prove (\ref{eq.lem.p1-1}a). Based on (\ref{eq.kkt-nonfading}d), we have $(\ell_i^{**})^2 = \frac{\eta h \tau^2 ( \bar{\theta}_i^{**} + \nu_i )}{3\zeta C^3}$, $\forall i\in{\cal N}$. By further considering the complementary slackness conditions of $\bar{\theta}_i^{**}\ell_i^{**}=0$ and $\bar{\theta}_i^{**}\geq 0$, it holds that
\begin{align}\label{eq.localKKT2}
&\bar{\theta}_i^{**} = 0 \Longrightarrow (\ell_i^{**})^2 = \frac{\eta h \tau^2 \nu_i }{3\zeta C^3} ~~{\rm and}~~ \bar{\theta}_i^{**} > 0 \Longrightarrow \ell_i^{**} = 0,~~\forall i\in{\cal N}.
\end{align}
From \eqref{eq.localKKT2}, we finally obtain (\ref{eq.lem.p1-1}a), where the operation $[\cdot]^+$ is implemented to guarantee $\ell_i^{**}\geq 0$.

Next, we prove (\ref{eq.lem.p1-1}b). Based on (\ref{eq.kkt-nonfading}c), it follows that $2^{\frac{d_i^{**}}{\tau B}} = (\underline{\theta}_i^{**} + \nu_i )\Big/ \Big(\frac{\sigma^2\ln2}{B\eta h g}\Big)$, $\forall i\in{\cal N}$. For each $i\in{\cal N}$, by considering the complementary slackness conditions of $\underline{\theta}_i^{**}d_i^{**}=0$ and $\underline{\theta}_i^{**}\geq 0$, we then have
\begin{align}\label{eq.localKKT3}
&\underline{\theta}_i^{**} = 0 \Longrightarrow 2^{\frac{d_i^{**}}{\tau B}} = \nu_i \Big/ \Big(\frac{\sigma^2\ln2}{B\eta h g}\Big) ~~{\rm and}~~ \underline{\theta}_i^{**} > 0 \Longrightarrow d_i^{**} = 0.
\end{align}
Therefore, based on \eqref{eq.localKKT3} and considering $d_i^{**}\geq 0$, the optimal number of task input-bits for offloading is obtained in (\ref{eq.lem.p1-1}b). Therefore, Theorem~1 is finally proved.
\end{IEEEproof}

For ease of description, we refer to $\nu_i$ in \eqref{eq.lem.p1-1} as the {\em computation level} at slot $i\in{\cal N}$. Based on the KKT optimality conditions in \eqref{eq.kkt-nonfading}, it is verified that the computation level is always nonnegative, i.e., $\nu_i\geq 0$, $\forall i\in{\cal N}$. In addition, since $\mu_j^{**}\geq 0$ and $\nu_{j+1}=\nu_j+\mu_j^{**}$, $\forall j\in{\cal N}\setminus\{N\}$, the computation level $\nu_i$ increases monotonically over slots, i.e., $\nu_1\leq \ldots\leq \nu_N$. Furthermore, we define $\nu_{N+1}\triangleq \infty$ and refer to slot $i\in{\cal N}$ as a {\em transition slot} if the computation level $\nu_i$ increases strictly after this slot, i.e., $\nu_i < \nu_{i+1}$. It is clear that the last slot $N$ is always a transition slot. Let ${\cal N}_{\rm TS}\triangleq \{\pi_1,\ldots, \pi_{|{\cal N}_{\rm TS}|}\}$ collect all the transition slots within the horizon such that $\pi_i < \pi_j$ for $i<j$ and $\pi_{|{\cal N}_{\rm TS}|}=N$. Based on Theorem 1 and the monotonically increasing nature of computation levels $\{\nu_i\}$ over time, we establish the following proposition.

\begin{proposition}\label{prop.mono}
The optimal task allocation of $\{\ell_i^{**}\}$ and $\{d_i^{**}\}$ for problem (${\cal P}2.1$) satisfies the so-called {\em staircase} property below.
\begin{itemize}
\item The number of task input-bits $\{\ell_i^{**}\}$ for local computing and $\{ d_i^{**}\}$ for offloading both increase monotonically over slots, i.e., $\ell_1^{**}\leq \ldots \leq \ell_N^{**}$ and $d_1^{**}\leq \ldots \leq d_N^{**}$.
\item If slot $i\in{\cal N}$ is a transition slot, then it holds that $\sum_{j=1}^i (\ell_j^{**}+d_j^{**}) = \sum_{j=1}^i A_j$, i.e., the task buffer at the user is completely cleared after this slot.
\end{itemize}
\end{proposition}

\begin{IEEEproof}
Based on \eqref{eq.lem.p1-1}, since $\nu_1\leq\ldots\leq \nu_N$, it yields that $\ell^{**}_1\leq \ldots\leq \ell^{**}_N$ and $d^{**}_1\leq \ldots\leq d^{**}_N$. Therefore, the first property of Proposition~\ref{prop.mono} is proved.

To prove the second property of Proposition~\ref{prop.mono}, we consider the cases of $i=N$ and $i\in{\cal N}\setminus\{N\}$, respectively. First, since all the cumulative tasks should be successfully computed before the end of the horizon, it must hold that $\sum_{j=1}^N(\ell_j^{**}+d_j^{**})=\sum_{j=1}^N A_j$. Next, we consider one particular slot $i\in{\cal N}\setminus\{N\}$ that is a transition slot, i.e., the computation level $\nu_i$ increases strictly after slot~$i$ with $\nu_i < \nu_{i+1}$. Given $\nu_i = -\sum_{j=i}^N \mu_j^{**}$ and $\nu_{i+1} = -\sum_{j=i+1}^N \mu_j^{**}$, we have $\mu^{**}_i>0$. Based on the complementary slackness conditions in (\ref{eq.kkt-nonfading}c), it follows that $\sum_{j=1}^i(\ell_j^{**}+d_j^{**})=\sum_{j=1}^i A_j$. The second property of Proposition~\ref{prop.mono} is thus verified.
\end{IEEEproof}

Notice that the staircase task allocation in Proposition~1 is reminiscent of the staircase energy allocation for energy harvesting powered wireless communications\cite{Rui12}. Motivated by \cite{Rui12}, we employ a {\em forward-search} procedure to find the optimal transition slot set, denoted by ${\cal N}^{**}_{\rm TS}=\{\pi_1^{**},\ldots,\pi^{**}_{|{\cal N}^{**}_{\rm TS}|}\}$, and then obtain the optimal task allocation $\{\ell_i^{**},d_i^{**}\}$ for problem (${\cal P}2.1$) (or equivalently (${\cal P}2$)), as presented as Algorithm~1 in Table~II and explained in detail as follows.

\begin{table}[htp]
\begin{center}
\caption{Algorithm 1 for Optimally Solving Problem (${\cal P}2.1$)}
\hrule
\begin{itemize}
\item[a)] {\bf Input:} The number of slots $N$, the task arrivals $\{A_i\}$, and the channel power gain $h$ for WPT and $g$ for offloading.
\item[b)]{\bf Initialize:} $\pi^{**}_0=0$.
\item[c)] {\bf For} $k=1,\ldots,N$ {\bf do}
\item[]\quad Define ${\cal N}_{\pi_k}^{\rm cand} \triangleq \{\pi^{**}_{k-1}+1,\ldots,N\}$ and set $\alpha_i \gets \frac{1}{i-\pi^{**}_{k-1}}{\sum_{j=\pi^{**}_{k-1}+1}^{i} A_j}$, $\forall i \in {\cal N}_{\pi_k}^{\rm cand}$;
\item[]\quad Set $\pi^{**}_k \gets \argmin_{i\in {\cal N}_{\pi_k}^{\rm cand} } \alpha_i$;
\item[] \quad Set $\ell_i^{**} \gets \tau\sqrt{\frac{[\nu_k]^+}{3\zeta C^3}}$, $d_i^{**} \gets \tau B\log_2\Big( \max\Big[ \nu_i \Big/ \Big(\frac{\sigma^2 \ln2} {B \eta h g}\Big),~1\Big] \Big)$, $\forall i \in \{\pi^{**}_{k-1}+1,\ldots,\pi^{**}_k\}$, where $\nu_k$ satisfies $\ell_i^{**}+d_i^{**} = \frac{1}{\pi^{**}_k - \pi^{**}_{k-1}} {\sum_{j=\pi^{**}_{k-1}+1}^{\pi^{**}_k} A_j}$;
\item[]\quad {\bf If} $\pi^{**}_k=N$ {\bf then}
\item[]\quad\quad Break;
\item[]\quad {\bf End if}
\item[]{\bf End for}
\item[c)] {\bf Output:} The optimal solution of $\{ \ell_i^{**},d_i^{**}\}$ for problem (${\cal P}2.1$).
\end{itemize}
\hrule
\end{center}
\end{table}

Algorithm 1 is implemented by induction, in which we start by searching the first optimal transition slot $\pi^{**}_1$, followed by $\pi^{**}_2$, $\pi^{**}_3$, $\ldots$, until the last optimal transition slot $\pi^{**}_{|{\cal N}^{**}_{\rm TS}|}=N$. In particular, the search of the $k$-th optimal transition slot $\pi^{**}_k$ is stated as follows. We define $\pi_0^{**}\triangleq 0$ for convenience. First, let ${\cal N}^{\rm cand}_{\pi_k}\triangleq \{\pi^{**}_{k-1}+1,\ldots,N\}$ denote the set of candidate transition slots. Then, for each candidate transition slot $i\in{\cal N}^{\rm cand}_{\pi_k}$, we compute $\ell_j+d_j=\frac{1}{i - \pi^{**}_{k-1}}{\sum^{i}_{m=\pi^{**}_{k-1}+1} A_m}$ as the unchanged number of task input-bits executed per-slot over slots $j\in\{\pi^{**}_{k-1}+1,\ldots,i\}$. Next, we choose $\pi^{**}_k = \argmin_{i\in{\cal N}_{\pi_k}^{\rm cand}} \frac{1}{i - \pi^{**}_{k-1}} {\sum^{i}_{j=\pi^{**}_{k-1}+1} A_j}$ as the $k$-th optimal transition slot, since this slot admits the smallest unchanged number of task input-bits per slot among all candidate slots in set ${\cal N}^{\rm cand}_{\pi_k}$. Given the optimal transition slot $\pi^{**}_k$ obtained, we have
\begin{align}\label{eq.ell}
\ell_i^{**}+d_i^{**} =\frac{1}{\pi^{**}_k - \pi^{**}_{k-1}}{\sum_{j=\pi^{**}_{k-1}+1}^{\pi^{**}_k} A_j},~~\forall i\in\{\pi^{**}_{k-1}+1,\ldots,\pi^{**}_k\},
\end{align}
where $\ell_i^{**}$ and $d_i^{**}$ are given in (\ref{eq.lem.p1-1}a) and (\ref{eq.lem.p1-1}b), respectively. Accordingly, we can find $\nu_i$ via a bisection search based on \eqref{eq.ell} and consequently find $\ell_i^{**}$ and $d_i^{**}$. Therefore, by performing the above procedures iteratively, the optimal task allocation of $\{\ell_i^{**}\}$ and $\{d_i^{**}\}$ is finally obtained. Note that the task allocation obtained in Algorithm~1 always satisfies the staircase property in Proposition~\ref{prop.mono} and equivalently the KKT conditions in \eqref{eq.kkt-nonfading}. Therefore, Algorithm~1 is ensured to achieve the optimal solution to problem (${\cal P}2.1$) and thus problem (${\cal P}2$).

\begin{remark}
In this paper, we develop Algorithm 1 by leveraging the monotonically increasing structure of the optimal number of task input-bits for offloading and local computing. In Algorithm 1, the iteration number at the worst case is equal to the slot number $N$. For each iteration $k\in{\cal N}$, there exists a candidate transition slot set ${\cal N}_{\pi_{k-1}}^{\rm cand}$, where $|{\cal N}_{\pi_{k-1}}^{\rm cand}|\leq N-k$. We only need to calculate the average number of task input-bits within the slot interval which starts from the previous transition slot $\pi_{k-1}^{**}$ to each candidate transition slot, and then choose the candidate slot with the smallest average number of task input-bits as the current transition slot $\pi_{k}^{**}$. By further solving \eqref{eq.ell}, we finally obtain the computation level and thus the optimal number of task input-bits for offloading and local computing, respectively. Therefore, at the worst case, Algorithm 1 involves a total of $\frac{N(N+1)}{2}$ addition operations and needs to implement \eqref{eq.ell} via bisection search for $N$ times.
\end{remark}

\subsection{Obtaining Optimal Energy Allocation $\{p_i^{**}\}$ to Problem (${\cal P}2$)}
Now, under the optimal task allocation $\{\ell_i^{**},d_i^{**}\}$ obtained by Algorithm 1, it remains to find the optimal energy allocation $\{p_i^{**}\}$ to problem (${\cal P}2$) based on (\ref{eq.prob2}b) and \eqref{eq.N}. Notice that based on Proposition~1, the allocated number of task input-bits and thus energy consumption at the user (for local computing and offloading) both monotonically increase over time. As a result, based on (\ref{eq.prob2}b) and \eqref{eq.N}, one optimal energy allocation solution to problem (${\cal P}2$) is to uniformly allocate energy for WPT over time by setting
\begin{align}\label{eq.energy-uniform}
p^{**}_i= \frac{1}{\tau \eta h N}\sum_{j=1}^N(E^{\rm loc}(\ell^{**}_j)+E^{\rm offl}(d^{**}_j)),~~\forall i\in{\cal N}.
\end{align}

By combining Algorithm~1 and \eqref{eq.energy-uniform}, we finally obtain the optimal offline solution $\{p_i^{**},\ell^{**}_i,d_i^{**}\}$ to problem (${\cal P}2$).

\begin{figure}
  \centering
  \begin{minipage}{8.1cm}
  \includegraphics[width = 2.9in]{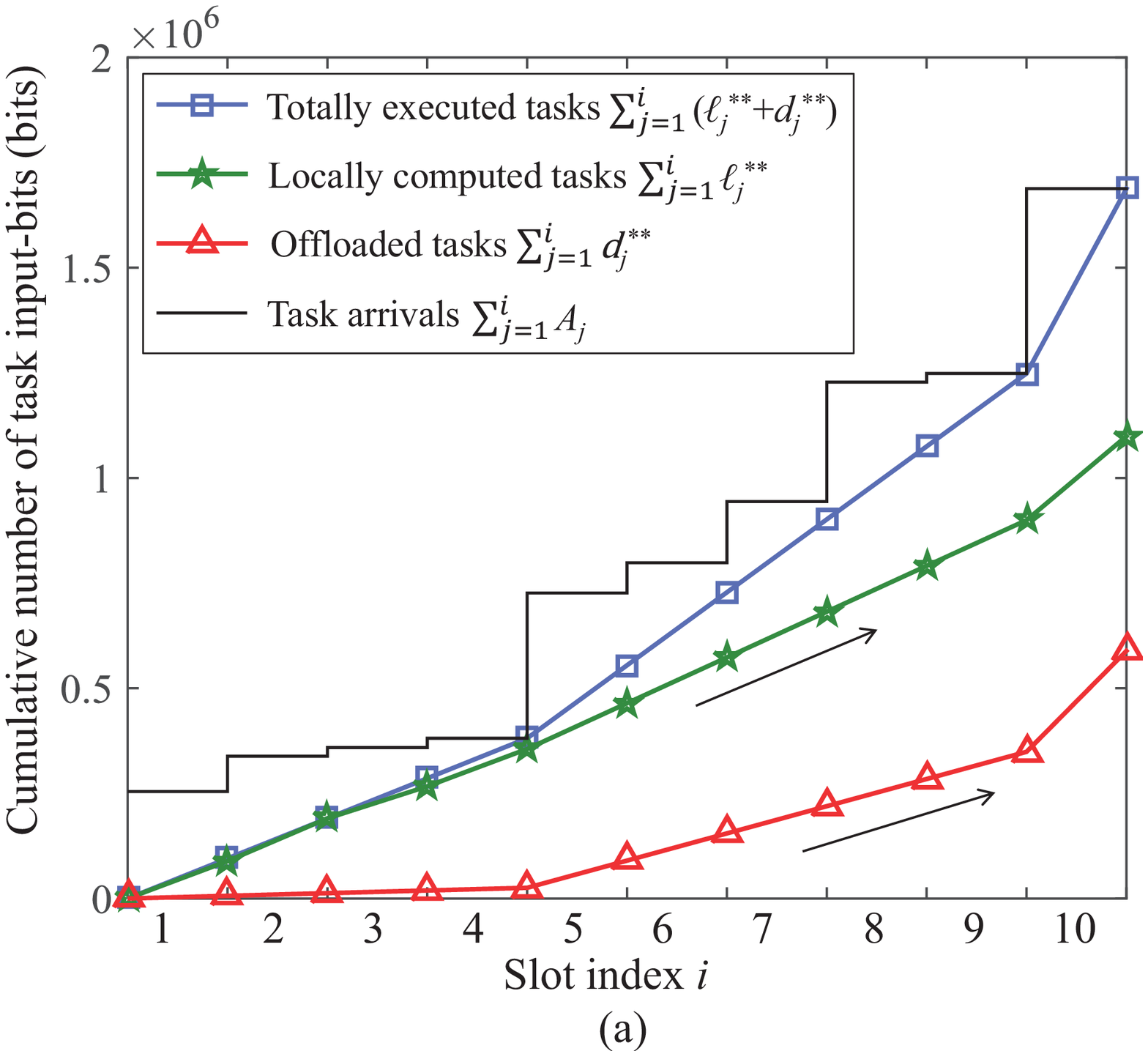}
\end{minipage}
\begin{minipage}{8.1cm}
  \includegraphics[width = 2.8in]{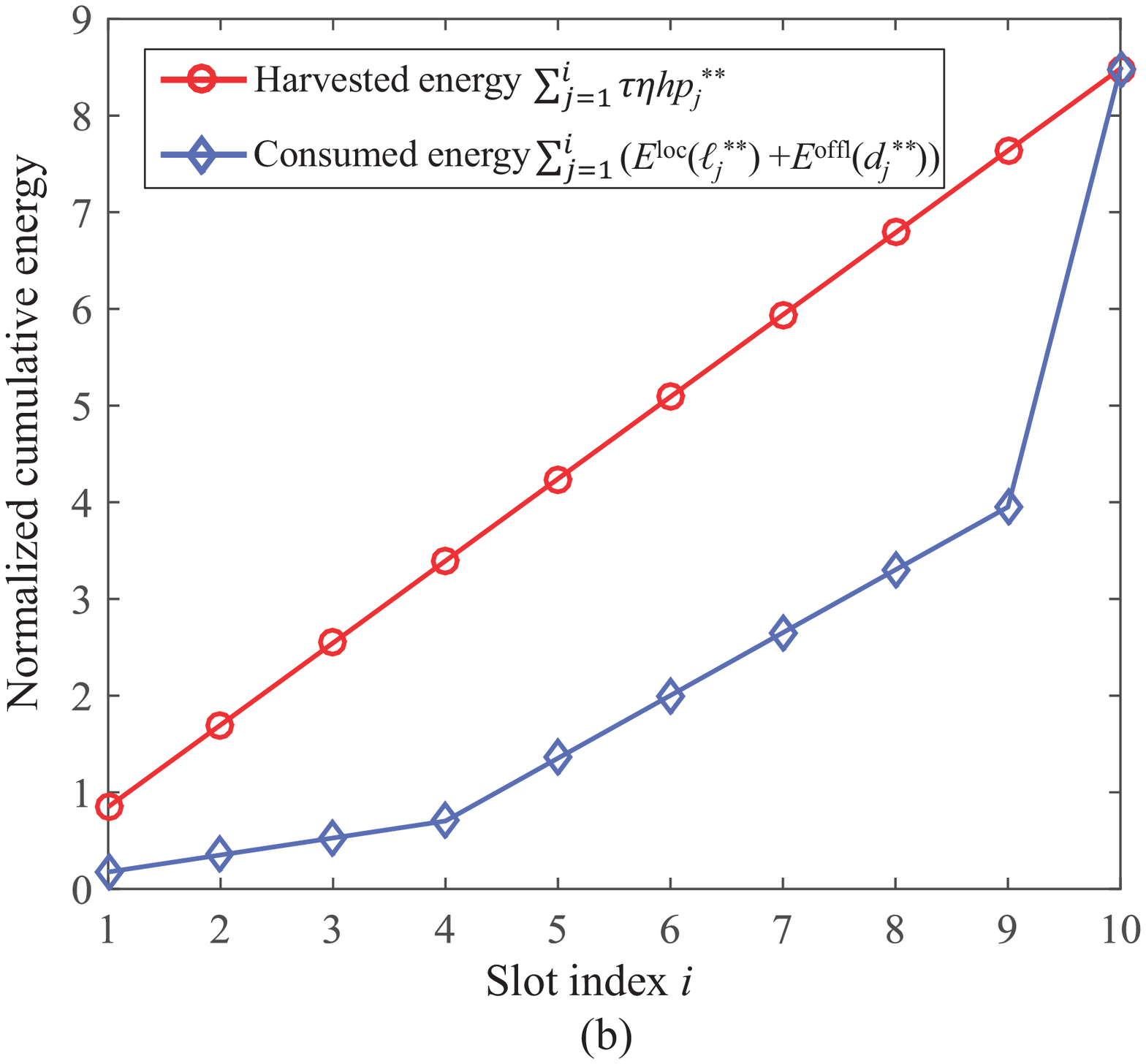}
  \end{minipage}
 \caption{An illustration of dynamic task arrivals $\{A_i\}$ and the optimal solution $\{p_i^{**},\ell_i^{**},d_i^{**}\}$ to problem (${\cal P}2$).} \label{fig.sta-solution-policy}
\end{figure}

\begin{example}
For illustration, Fig.~\ref{fig.sta-solution-policy} shows the optimal offline solution $\{\ell_i^{**},d_i^{**},p_i^{**}\}$ to problem (${\cal P}2$) with dynamic task arrivals $\{A_i\}$, where the number of slots is set to be $N=10$ and other system parameters are set same as those in Section~VI. As shown in Fig.~\ref{fig.sta-solution-policy}(a), there are in total $|{\cal N}^{**}_{\rm TS}|=3$ transition slots (i.e., $\pi^{**}_1=4$, $\pi^{**}_2=9$, and $\pi^{**}_3=10$), and the user's task buffer becomes empty after each of these transition slots. It is also observed that both $\{\ell^{**}_i\}$ and $\{d_i^{**}\}$ increase monotonically over time, and they remain unchanged within the corresponding transition slot intervals (i.e., $\{1,2,3,4\}$, $\{5,6,7,8,9\}$, and $\{10\}$). The observations in Fig.~\ref{fig.sta-solution-policy}(a) vividly corroborate the staircase task allocation structure (for local computing and offloading) as stated in Proposition~1. As shown in Fig.~\ref{fig.sta-solution-policy}(b), the proposed uniform energy allocation for WPT at the ET is easy to implement in practice for meeting the user's monotonically increasing computation energy demands over time.
\end{example}

\section{Optimal Energy and Task Allocation Under Time-Varying Channels}\label{sec:fading}
In this section, we present the optimal solution to problem (${\cal P}1$) under the general scenario with time-varying channels, where the channel power gains $\{h_i\}$ for WPT and $\{g_i\}$ for offloading may change over slots. Let $\{p_i^{*},\ell_i^{*},d_i^{*}\}$ denote the optimal solution to problem (${\cal P}1$).

\subsection{Decomposition of Problem (${\cal P}1$)}\label{sec:prob-decomp}
 In this subsection, we decouple problem (${\cal P}1$) into two subproblems for optimizing energy allocation $\{p_i\}$ and task allocation $\{\ell_i,d_i\}$, respectively. To this end, we first define the set of causality dominating slots (CDSs) for WPT from the ET to the user as \cite{XunZhou16}
\begin{align}\label{def.CDS}
{\cal N}_{\rm CDS} & \triangleq \{1\} \cup \Big\{ i\in \{2,\ldots, N\}\big|h_i > h_j, \forall 1\leq j < i\Big\} \notag \\
&= \{\phi_1, \ldots,\phi_{|{\cal N}_{\rm CDS}|}\},
\end{align}
where $1=\phi_1<\ldots<\phi_{|{\cal N}_{\rm CDS}|}\leq N$. It is clear that in set ${\cal N}_{\rm CDS}$, the channel power gain $h_{\phi_k}$ for WPT is strictly increasing over the CDS index $\phi_k$, i.e., $h_{\phi_1}<\ldots<h_{\phi_{|{\cal N}_{\rm CDS}|}}$. Then, we have the following theorem.

\begin{theorem}\label{lem.power_allocation}
Under any given task allocation of $\{\ell_i\}$ and $\{d_i\}$ at the user, the optimal energy allocation $\{p_i\}$ to problem (${\cal P}1$) is given by
\begin{align} \label{eq.opt_pi}
p_i =
\begin{cases}
 \frac{1}{\tau\eta h_{\phi_k}} \sum_{j=\phi_k}^{\phi_{k+1}-1} \left( E^{\rm loc}(\ell_j)+E_j^{\rm offl}(d_j)\right) ,~~& {\rm if}~ i=\phi_{k},~k\in\{1,\ldots,|{\cal N}_{\rm CDS}|\},\\
0,  &{\rm if}~ i\in{\cal N}\setminus{\cal N}_{\rm CDS},
\end{cases}
\end{align}
where $\phi_{|{\cal N}_{\rm CDS}|+1}\triangleq N+1$ is defined for convenience.
\end{theorem}
\begin{IEEEproof}
See Appendix \ref{proof_lem.power_allocation}.
\end{IEEEproof}
\begin{remark}
Theorem~\ref{lem.power_allocation} reveals the following two essential insights on the optimal energy allocation for WPT at the ET in wireless powered MEC systems over time-varying channels.
\begin{itemize}
\item
First, in order to the meet the energy demand at the user within the horizon, the ET should transmit wireless energy to the user only at CDSs, i.e., $p_i\neq 0$, $\forall i\in{\cal N}_{\rm CDS}$, and $p_j=0$, $\forall j\in{\cal N}\setminus{\cal N}_{\rm CDS}$. This is intuitively expected, since the user can always harvest a larger amount of energy when the ET allocates energy to an earlier CDS in ${\cal N}_{\rm CDS}$ rather than to other non-CDSs in ${\cal N}\setminus {\cal N}_{\rm CDS}$.

\item
Second, the amount of energy harvested by the user at each CDS $\phi_k$ equals that consumed by the user at the CDS interval $\{\phi_k,\ldots,\phi_{k+1}-1\}$, i.e., $\tau\eta h_{\phi_k}p_{\phi_k} = \sum_{j=\phi_k}^{\phi_{k+1}-1} \big( E^{\rm loc}(\ell_j)+E_j^{\rm offl}(d_j) \big)$, $\forall k\in\{1,\ldots,|{\cal N}_{\rm CDS}|\}$. This is because the channel power gains for WPT at CDSs are strictly increasing over time, and thus the ET only needs to allocate the exact amount of energy at the current CDS to meet the user's energy demand during the corresponding CDS interval.
\end{itemize}
\end{remark}

Based on Theorem~\ref{lem.power_allocation} and to facilitate the description, we define the {\em effective} channel power gain at slot $i\in\{\phi_k,\ldots,\phi_{k+1}-1\}$ for WPT from the ET to the user as
\begin{align}
h^\prime_i \triangleq h_{\phi_k},
\end{align}
where $k\in\{1,\ldots,|{\cal N}_{\rm CDS}|\}$. By substituting the optimal $p_i$'s (cf. \eqref{eq.opt_pi}) back into the objective function of problem (${\cal P}1$), it yields that
\begin{align}\label{eq.power-chain}
 \sum_{i=1}^N \tau p_i
&=\sum_{k=1}^{|{\cal N}_{\rm CDS}|} \sum_{i=\phi_k}^{\phi_{k+1}-1} \frac{1}{ \eta h_{\phi_{k}}} \left( E^{\rm loc}(\ell_i)+E_i^{\rm offl}(d_i) \right) = \sum_{i=1}^N \frac{1}{ \eta h_i^\prime } \left( E^{\rm loc}(\ell_i)+E_i^{\rm offl}(d_i) \right).
\end{align}
Based on \eqref{eq.power-chain}, we can obtain the optimal task allocation solution of $\{\ell^{*}_i\}$ and $\{d_i^{*}\}$ to problem (${\cal P}1$) by solving the following weighted sum energy minimization problem:
\begin{align*}\label{eq.prob01}
({\cal P}1.1):~~&\min_{\{\ell_{i}\geq 0,d_{i}\geq 0\}}
\sum_{i=1}^N \frac{1}{ \eta h_i^\prime } \left( E^{\rm loc}(\ell_i)+E_i^{\rm offl}(d_i) \right)  \\
& \quad ~~{\rm s.t.}\quad
 \eqref{eq.task-causality}~{\rm and}~\eqref{eq.task-completion}.
\end{align*}
In the following, we first derive the optimal task allocation solution of $\{\ell_i^*\}$ and $\{d_i^*\}$ at the user by solving problem (${\cal P}1.1$) and then obtain the optimal energy allocation $\{p_i^*\}$ at the ET to problem (${\cal P}1$) by using Theorem~\ref{lem.power_allocation}.

\subsection{Obtaining Optimal Task Allocation $\{\ell^{*}_i,d_i^{*}\}$ by Solving Problem (${\cal P}1.1$)}
As problem (${\cal P}1.1$) is a convex optimization problem that satisfies the Slater's condition, strong duality holds between problem (${\cal P}1.1$) and its Lagrange dual problem. Let $\lambda_i\geq 0$, $\forall i\in{\cal N}\setminus\{N\}$, $\lambda_N\in\mathbb{R}$, $\bar{\delta}_j\geq 0$, and $\underline{\delta}_j\geq 0$, $\forall j\in{\cal N}$, denote the Lagrange multipliers associated with the constraints in \eqref{eq.task-causality} and \eqref{eq.task-completion}, $\ell_j\geq 0$, and $d_j\geq 0$, respectively. The following KKT conditions are necessary and sufficient for $\{\ell_i^{*},d^{*}_i\}$ and $\{\lambda_i^{*},\bar{\delta}_i^{*},\underline{\delta}_i^{*}\}$ to be the primal and dual optimal solutions to problem~(${\cal P}1.1$)\cite{Boyd2004}.
\begin{subequations}\label{eq.kkt}
\begin{align}
& \ell_i^* \geq 0, \; d^*_i\geq 0,\; \bar{\delta}_i^* \geq 0,\; \underline{\delta}_i^{*} \geq 0,~\forall i\in{\cal N},\; \lambda_j^{*} \geq 0, ~\forall j\in{\cal N}\setminus\{N\}\\
&\sum_{j=1}^i(\ell_j^{*}+d_j^{*})- \sum_{j=1}^iA_j\leq 0,~\forall i\in{\cal N}\setminus\{N\},~\sum_{j=1}^N (\ell_j^{*}+d_j^{*}) - \sum_{j=1}^N A_j=0\\
&\bar{\delta}_i^{*} \ell_i^{*} =0, \; \underline{\delta}_i^{*} d_i^{*} =0,\;\lambda^{*}_i\Big[ \sum_{j=1}^i(\ell^{*}_j+d^{*}_j) -\sum_{j=1}^i A_j\Big] = 0,~~\forall i\in{\cal N}\\
&\frac{3\zeta C^3(\ell_i^{*})^2}{\eta h_i^\prime \tau^2} + \sum_{j=i}^N\lambda_j^{*} - \bar{\delta}_i^{*}=0, ~~\forall i\in{\cal N} \\
&\frac{\sigma^2\ln2}{B \eta h_i^\prime g_i}2^{\frac{d_i^{*}}{\tau B}} + \sum_{j=i}^N\lambda_j^{*} -\underline{\delta}_i^{*} =0,~~\forall i\in {\cal N},
\end{align}
\end{subequations}
where (\ref{eq.kkt}a--b) denote the primal and dual feasible conditions, (\ref{eq.kkt}c) denotes the complementary slackness conditions, and (\ref{eq.kkt}d) and (\ref{eq.kkt}e) mean that the gradients of the associated Lagrangian with respect to $\ell_i$ and $d_i$ vanish at $\ell_i=\ell_i^{*}$ and $d_i=d_i^{*}$, $\forall i\in{\cal N}$, respectively. Based on the KKT optimality conditions in \eqref{eq.kkt} together with some algebraic manipulations, one can obtain the closed-form solution of $\{\ell_i^{*},d_i^{*}\}$ to problem (${\cal P}1.1$) in the following theorem.
\begin{theorem} \label{theorem.sol}
For problem (${\cal P}1.1$), the optimal number of task input-bits $\ell_i^{*}$ for local computing and $d_i^{*}$ for offloading are given by
\begin{subequations}\label{eq.sol-fading}
\begin{align}
\ell^{*}_i &= \tau\sqrt{\frac{\eta h_i^\prime [\omega_i]^+}{3\zeta C^3}}, ~~\forall i\in {\cal N},\\
d_i^{*} &= \tau B\log_2 \Big( \max\Big[ \omega_i \Big/ \Big(\frac{\sigma^2\ln2}{B\eta h_i^\prime g_i}\Big), ~1 \Big]\Big),~~\forall i \in {\cal N},
\end{align}
\end{subequations}
respectively, where $\omega_i \triangleq -\sum_{j=i}^N\lambda_j^{*}$, $\forall i\in{\cal N}$.
\end{theorem}
\begin{IEEEproof}
Based on (\ref{eq.kkt}a--d), the optimal number of task input-bits $\ell_i^{*}$ for local computing is obtained as in (\ref{eq.sol-fading}a), where $\omega_i = -\sum_{j=i}^N\lambda_j^{*}$, $\forall i\in{\cal N}$. Similarly, based on (\ref{eq.kkt}a--c) and (\ref{eq.kkt}e), the optimal number of task input-bits $d_i^{*}$ for offloading is obtained as in (\ref{eq.sol-fading}b), $\forall i\in{\cal N}$.
\end{IEEEproof}

Analogously to the scenario with static channels, we refer to $\omega_i$ as the computation level at slot $i\in{\cal N}$. Similarly, it is verified that the computation level $\omega_i$ is nonnegative and monotonically increasing over time, i.e., $0\leq \omega_1\leq \ldots \leq \omega_N$. Also, we define $\omega_{N+1}\triangleq \infty$ and refer to slot $i\in{\cal N}$ as a transition slot if the computation level $\omega_i$ increases strictly after the $i$-th slot, i.e., $\omega_i< \omega_{i+1}$. Therefore, the last slot $N$ of the horizon is always a transition slot. We collect all the transition slots as set ${\cal N}_{\rm TS}=\{\pi_1,\ldots,\pi_{|{\cal N}_{\rm TS}|}\}$, such that $\pi_i < \pi_j$ for $i< j$ and $\pi_{|{\cal N}_{\rm TS}|}=N$. Based on Theorem 3 together with the monotonically increasing computation levels $\{\omega_i\}$, we establish the following proposition.

\begin{proposition}\label{prop.nondecreasing}
At the optimality of problem (${\cal P}1$), the task allocation of $\{\ell_i^{*}\}$ for local computing and $\{d_i^{*}\}$ for offloading satisfy the following properties.
\begin{itemize}
\item The task allocation for local computing has the staircase structure, i.e., the optimal number of task input-bits $\ell_i^*$ for local computing increases monotonically over time ($\ell^*_1\leq \ldots \leq \ell^*_{N}$).
\item The task allocation for offloading has the staircase water-filling structure, i.e., if the channel-dependent coefficient $\frac{\sigma^2\ln2}{B\eta h_i^\prime g_i}$ is smaller than the computation level $\omega_i$, we have the optimal number of task input-bits $d_i^*>0$ for offloading; otherwise, we have $d_i^*=0$. Furthermore, the computation level $\omega_i$ increases monotonically over time ($\omega_1\leq \ldots \leq \omega_N$).
\item If slot $i\in{\cal N}$ is a transition slot, then it holds that $\sum_{j=1}^i(\ell_j^{*}+d_j^{*})=\sum_{j=1}^i A_j$, i.e., the task buffer is cleared after slot $i$.
\end{itemize}
\end{proposition}
\begin{IEEEproof}
At each slot $i\in{\cal N}$, since the optimal number of task input-bits $\ell_i^{*}$ for local computing in (\ref{eq.sol-fading}a) is a strictly increasing function with respect to both $\omega_i$ and $h_i^\prime$. As $\omega_1\leq \ldots,\leq \omega_N$ and $h_1^\prime \leq \ldots \leq h_N^\prime$, the first property of Proposition~\ref{prop.nondecreasing} must hold. Furthermore, the second property can be readily verified based on (\ref{eq.sol-fading}b).

To prove the third property of Proposition~\ref{prop.nondecreasing}, we consider the cases of $i=N$ and $i\in{\cal N}\setminus\{N\}$, respectively. First, since all the cumulative tasks at the user should be computed before the end of slot $N$, it follows that $\sum_{j=1}^N(\ell_j^{*}+d_j^{*})=\sum_{j=1}^N A_j$. Next, we consider one particular transition slot $i\in{\cal N}\setminus\{N\}$ with $\omega_i < \omega_{i+1}$. Since $\omega_i = -\sum_{j=i}^N \lambda_j^{*}$ and $\omega_{i+1} = -\sum_{j=i+1}^N \lambda_j^{*}=\omega_i+\lambda_i^*$, it must hold that $\lambda^{*}_i>0$. Based on the complementary slackness conditions in (\ref{eq.kkt}c), it then follows that $\sum_{j=1}^i(\ell_j^{*}+d_j^{*})=\sum_{j=1}^i A_j$. The third property of Proposition~\ref{prop.nondecreasing} is thus proved.
\end{IEEEproof}

Next, based on Proposition~\ref{prop.nondecreasing} and Theorem 3, we can solve problem (${\cal P}1.1$) optimally by first minimizing the user's energy consumption under given possible sets of transition slots and then searching over the transition slots to find the one with the smallest energy consumption. In particular, for any transition slot set ${\cal N}_{\rm TS}=\{\pi_1,\ldots, \pi_{|{\cal N}_{\rm TS}|}\}$, we compute the minimum weighted sum energy consumption of the $k$-th transition slot interval (i.e., slots $\{\pi_{k-1}+1,\ldots, \pi_k\}$) as $E(\pi_{k-1}+1,\pi_k)$ by solving the following problem:
\begin{subequations}\label{eq.prob02}
\begin{align}%
{\cal SP}(\pi_{k-1}+1,{\pi_{k}}):~~
E(\pi_{k-1}+1,\pi_k) \triangleq & \min_{\{\ell_{j}\geq 0,d_{j}\geq 0\}}
\sum_{j=\pi_{k-1}+1}^{\pi_k} \frac{1}{ \eta h_j^\prime } \left( E^{\rm loc}(\ell_j)+E_j^{\rm offl}(d_j) \right)  \\
& ~~\quad{\rm s.t.}~~  \sum_{j={\pi_{k-1}+1}}^{\pi_k} (\ell_j + d_j) =\sum_{j={\pi_{k-1}+1}}^{\pi_k} A_j,
\end{align}
\end{subequations}
where $k\in\{1,\ldots,|{\cal N}_{\rm TS}|\}$ and $\pi_0\triangleq 0$ is defined for convenience. It is worth noting that a new task completion constraint (\ref{eq.prob02}b) from slot $\pi_{k-1}+1$ to slot $\pi_k$ is imposed in problem ${\cal SP}(\pi_{k-1}+1,\pi_{k})$ due to the third property in Proposition~2, and meanwhile, the task causality constraints at slots $\{\pi_{k-1}+1,\ldots,\pi_k\}$ have been safely removed based on Theorem~3. At the optimality of problem (${\cal P}1.1$), the search of the optimal transition slot set, denoted by ${\cal N}_{\rm TS}^*=\{\pi_1^*,\ldots,\pi^*_{|{\cal N}^*_{\rm TS}|}\}$, is then formulated as
\begin{align}\label{eq.prob-search}
({\cal P}1.2):~~{\cal N}_{\rm TS}^* \triangleq \argmin_{{\cal N}_{\rm TS}\subseteq {\cal N}}
\sum_{k=1}^{|{\cal N}_{\rm TS}|} E(\pi_{k-1}+1,\pi_k).
\end{align}
Let $\{\ell_j^{(\pi_k)},d_j^{(\pi_k)}\}_{j=\pi_{k-1}+1}^{\pi_k}$ denote the optimal solution to problem ${\cal SP}(\pi_{k-1}+1,\pi_{k})$. Note that problem ${\cal SP}(\pi_{k-1}+1,\pi_{k})$ is convex and satisfies the Slater's condition. Based on the KKT optimality conditions for problem ${\cal SP}(\pi_{k-1}+1,\pi_{k})$, we can obtain the optimal solution as $\ell^{(\pi_k)}_j = \tau\sqrt{\frac{\eta h_j^\prime [\omega^{(\pi_k)}]^+}{3\zeta C^3}}$ and $d^{(\pi_k)}_j = \tau B\log_2 \Big(\max\Big[\omega^{(\pi_k)}\Big/ \Big(\frac{\sigma^2\ln2}{B\eta h_j^\prime g_j}\Big), ~1 \Big]\Big)$, $\forall j\in\{\pi_{k-1}+1,\ldots,\pi_k\}$, where the computation level $\omega^{(\pi_k)}$ is actually the optimal Lagrange multiplier associated with constraint (\ref{eq.prob02}b) that can be readily found via a bisection search based on the equality $\sum_{j={\pi_{k-1}+1}}^{\pi_k} (\ell^{(\pi_k)}_j + d^{(\pi_k)}_j) =\sum_{j={\pi_{k-1}+1}}^{\pi_k} A_j$.

Now, we solve problem (${\cal P}1.2$) to find the optimal transition slot set ${\cal N}_{\rm TS}^{*}$, and accordingly solve problem (${\cal P}1.1$). Similarly as in Section III and inspired by the staircase water-filling energy allocation in \cite{Rui12}, we employ a forward-search procedure to find the optimal ${\cal N}_{\rm TS}^{*}$ for problem (${\cal P}1.2$) and then obtain the optimal task allocation $\{\ell_i^{*},d_i^{*}\}$ for problem (${\cal P}1.1$), as presented as Algorithm 2 in Table~III and detailed below.

\begin{table}[htp]
\begin{center}
\caption{Algorithm 2 for Optimally Solving Problem (${\cal P}1.1$)}
\hrule
\begin{itemize}
\item[a)] {\bf Input:} The number of slots $N$, the task arrivals $\{A_i\}$, the effective channel power gains $\{h^\prime_i\}$ for WPT, and the channel power gains $\{g_i\}$ for offloading.
\item[b)] {\bf Initialization:} $\pi^{*}_0 = 0$ and ${\cal N}_{\rm TS}^* =\emptyset$.
\item[c)] {\bf For} $k=1,\ldots,N$ {\bf do}
\begin{itemize}
\item[] Set ${\cal N}^{\rm fea}_k \gets \emptyset$ and define ${\cal N}_{\pi_k}^{\rm cand} \triangleq \{\pi^*_{k-1}+1,\ldots,N\}$;
\item[] Obtain $\{\ell_j^{(i)},d_j^{(i)}\}_{j=\pi^*_{k-1}+1}^{i}$ by solving problem ${\cal SP}(\pi^{*}_{k-1}+1,i)$, $\forall i \in {\cal N}_{\pi_k}^{\rm cand}$;
\item[] Set ${\cal N}_k\gets {\cal N}_k\cup\{i\}$ if $\sum_{j=\pi^*_{k-1}+1}^m (\ell_j^{(i)}+d_j^{(i)}) \leq \sum_{j=\pi^*_{k-1}+1}^m A_j$, $\forall m\in \{\pi^{*}_{k-1}+1,\ldots,i\}, i\in{\cal N}_{\pi_k}^{\rm cand}$;
\item[] Obtain $\pi_k^*=\argmax_{\pi_k\in{\cal N}_k} \pi_k$;
\item[] Set ${\cal N}_{\rm TS}^* \gets {\cal N}_{\rm TS}^*\cup\{\pi_k^*\}$;
 \item[]  {\bf If} $\pi_k^{*}=N$ {\bf then}\\
\quad\quad Break;\\
\item[]{\bf End if}
\end{itemize}
\item[] {\bf End for}
\item[d)] {\bf Output:} The optimal $\ell_j^*=\ell_j^{(\pi^*_k)}$ and $d_j^*=d_j^{(\pi^*_k)}$, $\forall j\in\{\pi_{k-1}^*+1,\ldots,\pi_k^*\}$, $k\in\{1,\ldots,|{\cal N}^*_{\rm TS}|\}$, for problem (${\cal P}1.1$).
\end{itemize}
\hrule
\end{center}
\end{table}

Algorithm 2 is implemented by induction, in which we start from the search of the first optimal transition slot $\pi_1^{*}$, followed by $\pi_2^{*}$, $\pi_3^{*}$, $\ldots$, until the last optimal transition slot $\pi^{*}_{|{\cal N}^*_{\rm TS}|}=N$. We define $\pi_0^*\triangleq 0$ for convenience. In particular, the search of the $k$-th optimal transition slot $\pi^*_k$ is stated as follows. First, we let ${\cal N}^{\rm cand}_{\pi_k}\triangleq \{\pi^*_{k-1}+1,\ldots,N\}$ denote the set of candidate transition slots, and denote ${\cal N}^{\rm fea}_k$ as the set of feasible transition slots that is initialized as ${\cal N}_k^{\rm fea}\gets \emptyset$. Then, for each candidate transition slot $i\in{\cal N}^{\rm cand}_{\pi_k}$, we obtain the optimal task allocation of $\{\ell^{(i)}_j, d^{(i)}_j\}_{j=\pi^*_{k-1}+1}^{i}$ by solving problem ${\cal SP}(\pi^*_{k-1}+1,i)$. If the obtained $\{\ell^{(i)}_j, d^{(i)}_j\}_{j=\pi^*_{k-1}+1}^{i}$ satisfy the task causality constraints $\sum_{j=\pi^*_{k-1}+1}^m(\ell^{(i)}_j+d_j^{(i)})\leq \sum_{j={\pi^*_{k-1}+1}}^m A_j$, $\forall m\in\{ \pi^*_{k-1}+1,\ldots,i\}$, then we admit slot $i$ into set ${\cal N}^{\rm fea}_k$ by setting ${\cal N}^{\rm fea}_k = {\cal N}^{\rm fea}_k\cup\{i\}$. Finally, we choose the slot $\pi_k^* = \argmax_{\pi_k\in{\cal N}^{\rm fea}_k} \pi_k$ as the $k$-th optimal transition slot for problem (${\cal P}1.1$). Notice that the optimality of forward searching slots $\{\pi_k^*\}$ can be similarly verified based on the proof in \cite[Lemma 1]{Rui12}, for which the details are omitted for brevity. Therefore, by Algorithm 2, we finally find the optimal transition slot set ${\cal N}_{\rm TS}^*$ for problem (${\cal P}1.2$), and obtain the optimal task allocation of $\{\ell_j^{(\pi_k^*)},d_j^{(\pi_k^*)}\}_{j=\pi_{k-1}^*+1}^{\pi_k^*}$, $\forall k\in\{1,\ldots,|{\cal N}_{\rm TS}^*|\}$, for problem (${\cal P}1.1$).

\begin{figure}
  \centering
  \begin{minipage}{8.1cm}
  \includegraphics[width = 2.9in]{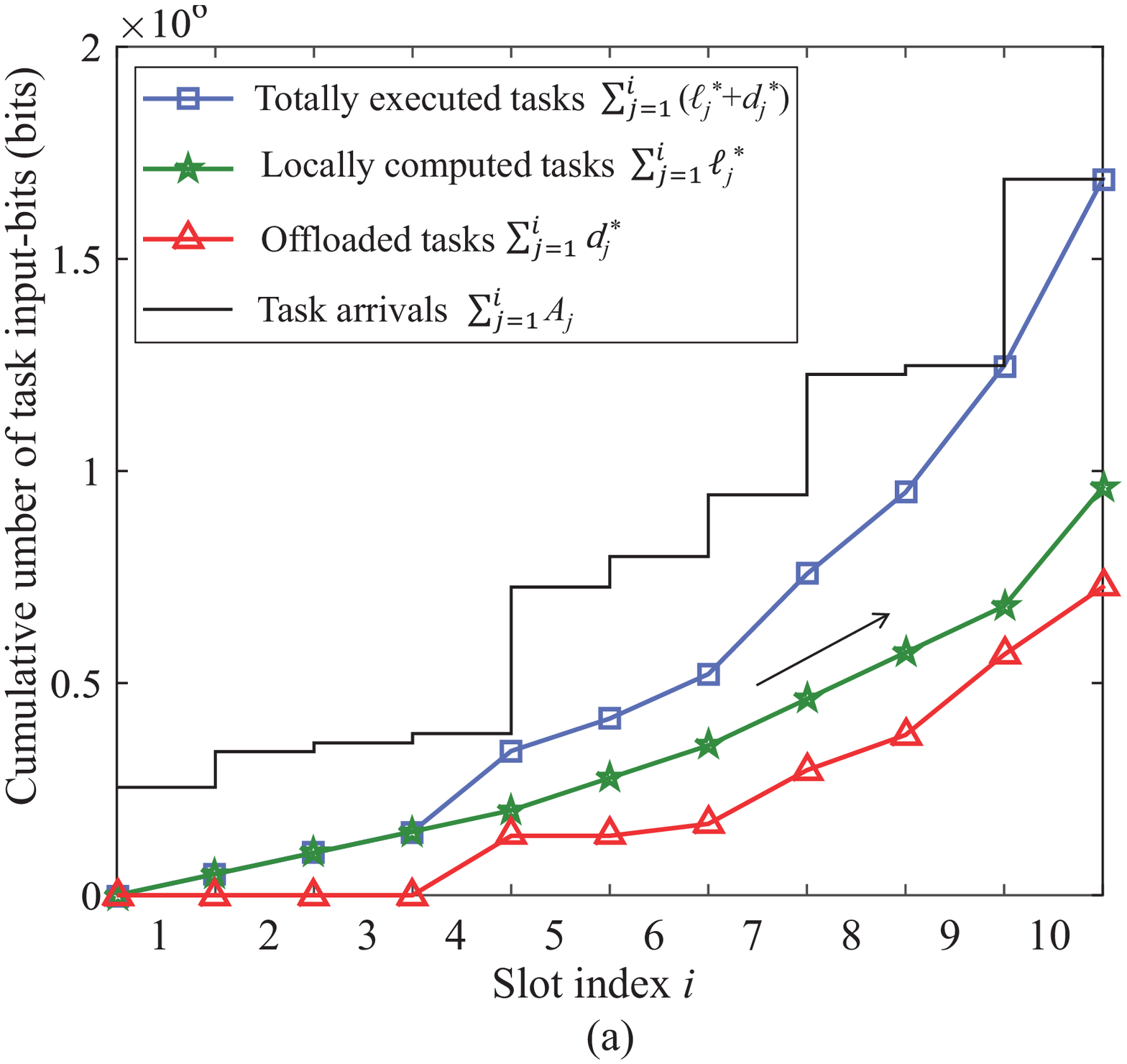}
  \end{minipage}
  \begin{minipage}{8.1cm}
  \centering
  \includegraphics[width = 2.8in]{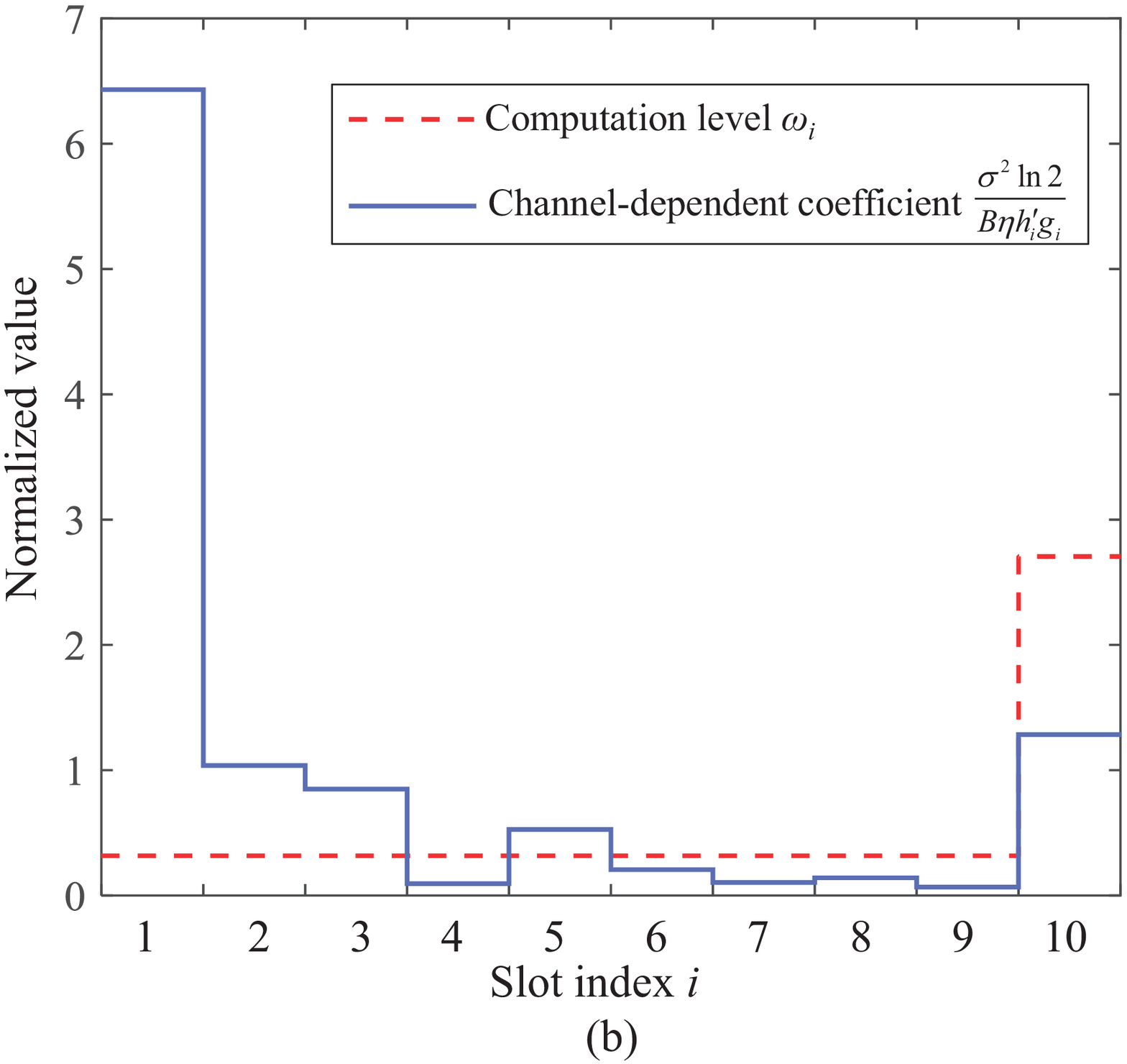}
  \end{minipage}
  \begin{minipage}{8.1cm}
  \includegraphics[width = 2.9in]{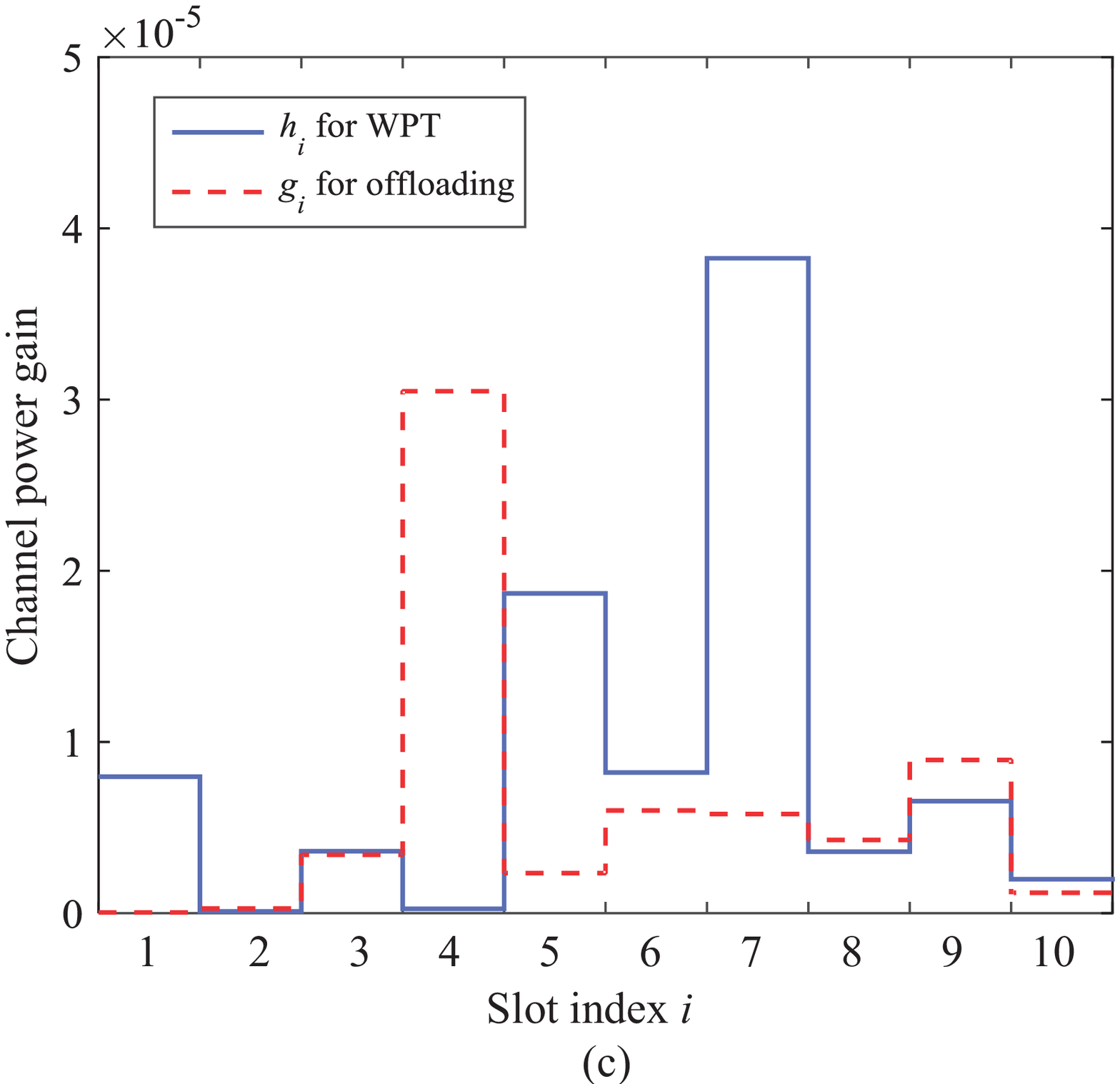}
  \end{minipage}
  \begin{minipage}{8.1cm}
  \centering
  \includegraphics[width = 2.9in]{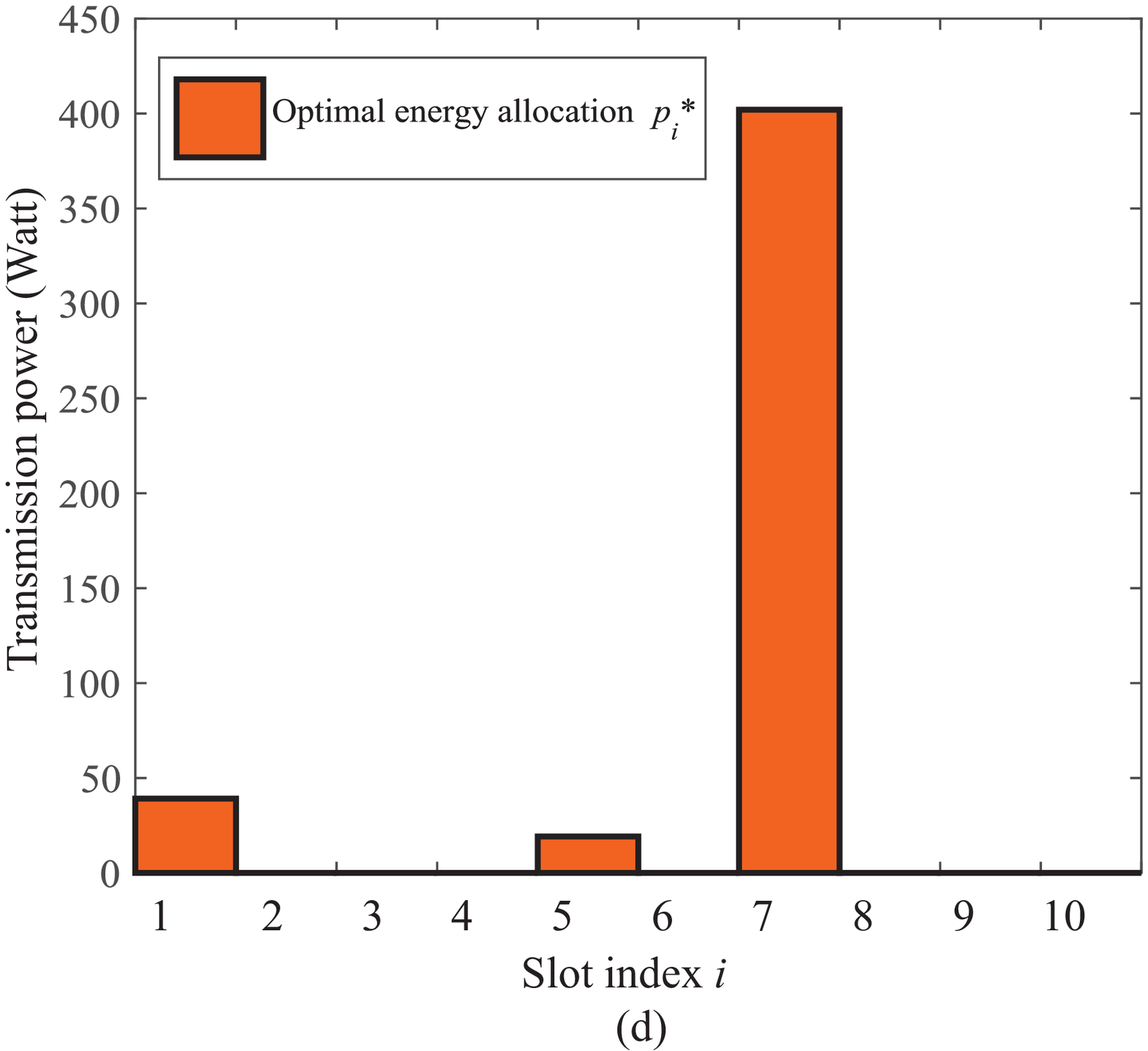}
  \end{minipage}
\caption{An illustration of dynamic task arrivals and the optimal offline solution $\{p_i^*,\ell_i^*,d_i^*\}$ to problem (${\cal P}1$).}\label{fig.Sol-dynamic}
\end{figure}

\subsection{Obtaining Optimal Energy Allocation Solution $\{p^{*}_i\}$ to Problem (${\cal P}1$)}

Based on Theorem 2, we can obtain the optimal energy allocation $\{p^*_i\}$ to problem (${\cal P}1$) based on \eqref{eq.opt_pi}, by replacing $\{\ell_i,d_i\}$ with the obtained $\{\ell_i^{*},d_i^{*}\}$ by Algorithm 2 above.

Until now, we finally obtain the optimal offline solution of $\{p_i^{*}, \ell_i^{*},d_i^{*}\}$ to problem (${\cal P}1$).

\begin{example}
For illustration, Fig.~\ref{fig.Sol-dynamic} shows the optimal offline solution to problem (${\cal P}1$) in the scenario with time-varying channels, where the number of slots is set to be $N=10$ and other system parameters are set same as those in Section VI. In Fig.~\ref{fig.Sol-dynamic}(a), we observe $|{\cal N}_{\rm TS}^*|=2$ optimal transition slots (i.e., $\pi_1^*=9$ and $\pi_2^*=10$), at which the user's task buffer becomes empty, and the task allocation of $\ell_i^*$ for local computing increases monotonically over slots. As shown in Fig.~\ref{fig.Sol-dynamic}(b), the computation level $\omega_i$ increases strictly after the first optimal transition slot $\pi_1^*=9$, and the channel-dependent coefficient $\frac{\sigma^2\ln2}{B\eta h_i^\prime g_i}$ is higher than the computation level $\omega_i$ at slots $\{1,2,3,5\}$, which is consistent with $d_i^*=0$ at these slots in Fig.~\ref{fig.Sol-dynamic}(a). It is also observed that a large ratio of the computation level to the channel-dependent coefficient leads to a large $d_i^*$ value (e.g., $d_9^*>d_4^*>d_6^*$). These observations in Figs.~\ref{fig.Sol-dynamic}(a) and~\ref{fig.Sol-dynamic}(b) vividly corroborate the staircase water-filling structure of task allocation $\{d^*_i\}$ for offloading in Proposition~2. Fig.~\ref{fig.Sol-dynamic}(c) shows the channel power gains $\{h_i\}$ for WPT and $\{g_i\}$ for offloading. It is observed that there are a total of $|{\cal N}_{\rm CDS}|=3$ CDSs (i.e., $\phi_1=1$, $\phi_2=5$, and $\phi_3=7$) for WPT. Fig.~\ref{fig.Sol-dynamic}(d) shows the transmission energy allocation for WPT at the ET. It is observed that the ET allocates energy only at these three CDSs, i.e., $p^*_i>0$ for $i\in\{1,5,7\}$ and $p^*_j=0$ for $j\in\{2,3,4,6,8,9,10\}$, which is consistent with Theorem~2.
\end{example}

\section{Heuristic Online Designs for Joint Energy and Task Allocation}
In the previous two sections, we have studied the offline optimization for the joint energy and task allocation in wireless powered MEC systems by assuming that the CSI/TSI is perfectly known {\em a-priori}. In this section, we consider that only the causal (i.e., the past and current) CSI and TSI are available. Inspired by the structures of the optimal offline solutions, we propose heuristic online designs in the scenarios with static and time-varying channels, respectively.

For the purpose of exposition, in this section we assume that the task arrivals $\{A_i\}$ at different slots follow a stochastic process with a given mean $A_{\rm mean}$. It is assumed that the ET/AP are able to obtain the value $A_{\rm mean}$, but they do not necessarily know the exact distribution of the stochastic process. For the scenario with time-varying channels, it is assumed that the channel power gains $\{h_i\}$ for WPT and $\{g_i\}$ for offloading within the horizon are generated based on a stochastic process with mean $\bar{h}$ and $\bar{g}$, respectively. The ET/AP knows the values of ${\bar h}$ and ${\bar g}$, but does not know the distribution of $\{h_i\}$ and $\{g_i\}$. Under this setup, the online joint energy and task allocation design is obtained at each slot $i\in{\cal N}$ subsequently, by minimizing the total energy consumption from the current slot $i$ to the last slot $N$, via viewing the mean values of channel power gains and task arrival amounts (known {\em a-priori}) as the estimated ones in future~slots.

\subsection{Static Channel Scenario}
First, we consider the scenario with static channels, in which the static CSI $h$ for WPT and $g$ for offloading are known. In this scenario, we denote $\{p_i^{\rm sta}\}$, $\{\ell_i^{\rm sta}\}$, and $\{d_i^{\rm sta}\}$ as the obtained online energy allocation, and task allocation for local computing and offloading, respectively. Let $R^{\rm sta}_i\triangleq \sum_{j=1}^{i} (A_j -\ell^{\rm sta}_j - d_j^{\rm sta})$ denote the residual number of task input-bits at the user's task buffer at the end of slot $i\in{\cal N}$ and we define $R^{\rm sta}_0\triangleq 0$.

Now, we consider any one particular slot $i\in{\cal N}$. In this slot, the online design is implemented by minimizing the total transmission energy consumption at the ET from slot $i$ to slot $N$, which can be formulated similarly as problem (${\cal P}2.1$) by regarding slot $i$ as the first slot, setting the total number of slots as $N-i+1$, and viewing $A_i+R_{i-1}^{\rm sta}$ as the exact amount of arrived tasks at slot $i$ and $A_{\rm mean}$ as the estimated amount of arrived tasks at each of subsequent slots in $\{i+1,\ldots,N\}$, respectively. Based on Theorem~1 and the staircase task allocation structure in Proposition~1, we can obtain the optimal solution to the energy minimization problem of interest as $\ell_j^{{\rm sta}-(i)}$ and $d_j^{{\rm sta}-(i)}$, $\forall j\in\{i,\ldots, N\}$. Accordingly, we set the task allocation of $\ell_i^{{\rm sta}-(i)}$ and $d_i^{{\rm sta}-(i)}$ at slot $i$ as the online design at this slot, i.e., $\ell_i^{\rm sta}=\ell_i^{{\rm sta}-(i)}$ and $d_i^{\rm sta}=d_i^{{\rm sta}-(i)}$. Then, under the obtained task allocation of $\ell_i^{\rm sta}$ and $d_i^{\rm sta}$ at slot $i\in{\cal N}$, we obtain the online energy allocation $p_i^{\rm sta}=\frac{1}{\tau \eta h}\big(E^{\rm loc}(\ell^{\rm sta}_i)+E^{\rm offl}(d^{\rm sta}_i)\big)$ at slot $i$, such that the wireless energy transferred to the user at slot $i$ equals that consumed by local computing and offloading at that slot.


\subsection{Time-varying Channel Scenario}
Next, we consider the scenario with time-varying channels. Let $\{p_i^{\rm tv}\}$, $\{\ell_i^{\rm tv}\}$, and $\{d_i^{\rm tv}\}$ denote the proposed online energy allocation, and task allocation for local computing and offloading, respectively. We denote $R^{\rm tv}_i\triangleq \sum_{j=1}^{i} (A_j -\ell^{\rm tv}_j - d_j^{\rm tv})$ as the residual number of task input-bits at the user's task buffer at the beginning of slot $i\in{\cal N}$ and define $R_0^{\rm tv}\triangleq 0$.

Now, we consider any one particular slot $i\in{\cal N}$. In this slot, we determine the effective channel power gains for WPT as $h_i^\prime = h_i$ and $h_j^\prime = \max(h_i,\bar{h})$, $\forall j\in\{i+1,\ldots,N\}$. Then, we consider the total transmission energy minimization from slot $i\in{\cal N}$ to slot $N$, which can be formulated similarly as problem (${\cal P}1.1$) by regarding slot $i$ as the first slot, setting the total number of slot as $N-i+1$, replacing $g_j$ with the estimated $\bar{g}$ at each slot $j\in\{i+1,\ldots,N\}$, and viewing $A_i+R_{i-1}^{\rm tv}$ as the exact amount of arrived tasks at slot $i$ and $A_{\rm mean}$ as the estimated amount of arrived tasks at slots $\{i+1,\ldots,N\}$, respectively. Based on Theorem~3, we have the optimal solution as $\ell_j^{{\rm tv}-(i)}$ and $d_j^{{\rm tv}-(i)}$, $\forall j\in\{i,\ldots,N\}$, which can then be efficiently computed by Algorithm 2 with slight parameter modification. Accordingly, we set the task allocation of $\ell_i^{{\rm tv}-(i)}$ and $d_i^{{\rm tv}-(i)}$ at slot $i$ as the online design at this slot, i.e., $\ell_i^{\rm tv}=\ell_i^{{\rm tv}-(i)}$ and $d_i^{\rm tv} = d_i^{{\rm tv}-(i)}$.

Then, we compute the online energy allocation $\{p_i^{\rm tv}\}$ at the ET under the $\ell_i^{\rm tv}$ and $d_i^{\rm tv}$ obtained above. To this end, we denote $S_i^{\rm tv} \triangleq \sum_{j=1}^{i} (\tau\eta h_jp_j^{\rm tv} - E^{\rm loc}(\ell_j^{\rm tv}) - E_j^{\rm offl}(d_j^{\rm tv}))$ as the residual energy amount at the user's energy storage at the end of slot $i\in{\cal N}$ and define $S^{\rm tv}_0 \triangleq 0$. For slot $i\in{\cal N}\setminus\{N\}$, we heuristically consider the following threshold based policy for energy allocation at the ET:
\begin{align}\label{eq.online_power}
p_i^{\rm tv} =
\begin{cases}
\frac{1}{\tau\eta h_i}\big[ E^{\rm loc}(\ell_i^{\rm tv}) + E_i^{\rm offl}(d_i^{\rm tv}) - S_{i-1}^{\rm tv} \big]^+,&{\rm if}~ h_i \leq \bar{h},\\
\frac{1}{\tau\eta h_i}\big[ \gamma \big(E^{\rm loc}(\ell_i^{\rm tv}) + E_i^{\rm offl}(d_i^{\rm tv})\big) - S_{i-1}^{\rm tv} \big]^+,&{\rm if}~ h_i>\bar{h},
\end{cases}
\end{align}
where $\gamma>1$ is a parameter for balancing energy supply and channel power gain at the current slot. The proposed energy allocation solution in \eqref{eq.online_power} indicates that if $h_i>\bar{h}$ (i.e., the channel condition is admirable for WPT), then the ET allocates more energy to exploit the large channel power gain; otherwise, the ET just allocates the minimum energy to meet the user's computation energy demand at that slot. In our design, we set $\gamma=2$. Furthermore, since there is no task to be executed after slot $N$, we have $p^{\rm tv}_N = \frac{1}{\tau \eta h_N}\big[ E^{\rm loc}(\ell_N^{\rm tv}) + E_N^{\rm offl}(d_N^{\rm tv}) - S^{\rm tv}_{N} \big]^+$. Until now, we finally obtain a heuristic online solution $\{p_i^{\rm tv},\ell_i^{\rm tv},d_i^{\rm tv}\}$ in the scenario with time-varying channels.

\section{Numerical Results}
In this section, we provide numerical results to evaluate the performance of the proposed designs. In the simulations, the system parameters are set as follows, unless stated otherwise. We set $M=4$, $\eta=0.3$, $C=200$ CPU cycles/bit, $\zeta=10^{-29}$\cite{Burd96}, the receiver noise power $\sigma^2=10^{-9}$~Watt, the slot length $\tau=0.1$ second, and the system bandwidth for offloading $B=1$ MHz. At slot $i\in{\cal N}$, the number of task input-bits is uniformly distributed as $A_i\sim {\cal U}[0,A_{\max}]$ with mean $A_{\rm mean}={A_{\max}}/{2}$. We consider that the ET and the AP are located with a distance of 10 meters (m) and the user is located on the line between them. Denote $d$ as the distance from the user to the ET. We consider the distance-dependent Rician fading channel models \cite{You16}, which are given as $\hat{\bm h}_i=\sqrt{\frac{{\cal X}_R \Omega_0 d^{-\kappa}}{1+{\cal X}_R}}{\bm h}_0 +\sqrt{\frac{\Omega_0 d^{-\kappa}}{1+{\cal X}_R}}{\tilde{\bm h}}$ and ${\hat g}_i=\sqrt{\frac{{\cal X}_R\Omega_0 (10-d)^{-\kappa}}{1+{\cal X}_R}}g_0+\sqrt{\frac{\Omega_0 (10-d)^{-\kappa}}{1+{\cal X}_R}}\tilde{g}$, $i\in{\cal N}$, respectively, where ${\cal X}_R=2$ is set as the Rician factor, $\Omega_0=-37$ dB corresponds to the path loss at a reference distance of one meter, $\kappa=3$ is the pathloss exponent, the line-of-sight (LoS) components ${\bm h}_0$ and $g_0$ have all elements equal to one, and the terms $\tilde{\bm h}\sim {\cal CN}({\bm 0},{\bm I})$ and $\tilde{g}\sim {\cal CN}(0,1)$ account for small-scale fading. The channel power gains for WPT and offloading are given by $h_i=\|\hat{\bm h}_i\|^2$ and $g_i=|\hat{g}_i|^2$, respectively. The numerical results are obtained by averaging over $10^3$ randomized channel and task realizations.

\begin{figure}
  \centering
  \begin{minipage}{8cm}
  \includegraphics[width = 8.0cm]{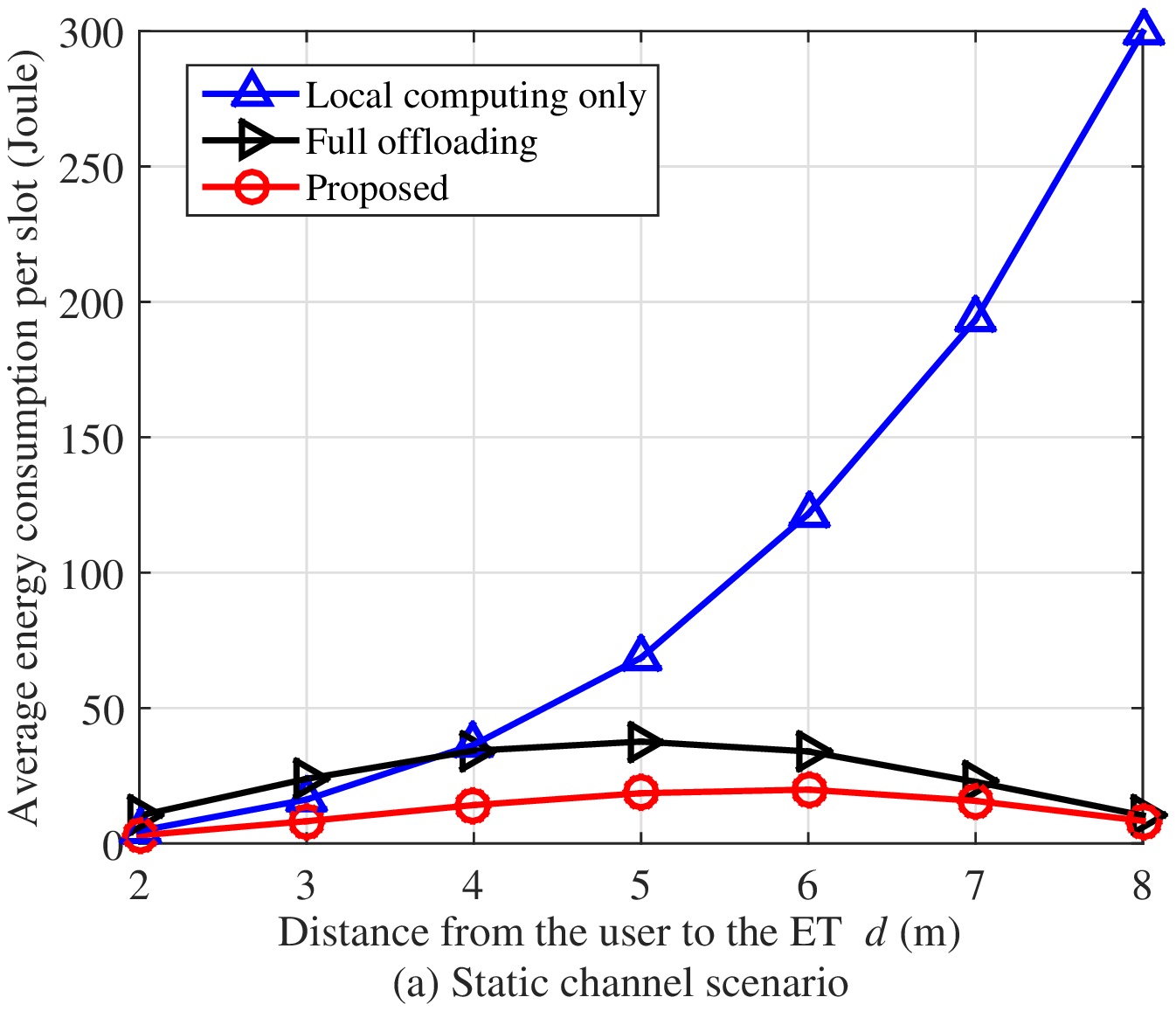}
  \end{minipage}
  \begin{minipage}{8cm}
  \includegraphics[width = 8.0cm]{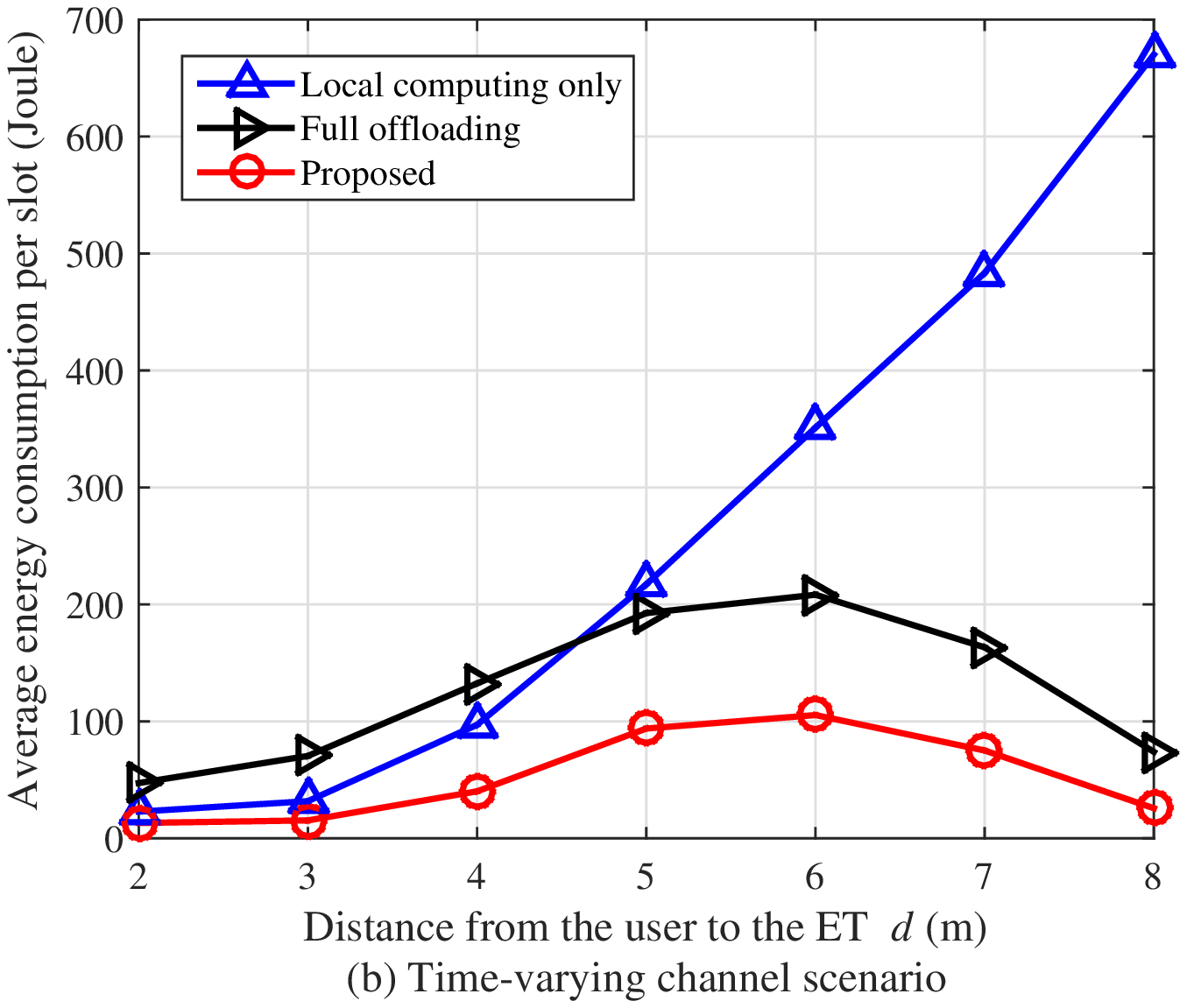}
  \end{minipage}
 \caption{Average energy consumption per slot at the ET versus the distance $d$ from the user to the ET.} \label{fig.vs-d}
\end{figure}

First, we consider the offline designs with the noncausal CSI/TSI known {\em a-priori}. Figs.~\ref{fig.vs-d}(a) and \ref{fig.vs-d}(b) show the average energy consumption at the ET per slot versus the distance $d$ from the user to the ET in the scenarios with static and time-varying channels, respectively, where $N=50$ and $A_{\max}=5\times 10^5$ bits. For comparison, we consider the local-computing-only and the full-offloading schemes as benchmark schemes, in which the user's computation tasks are executed only by local computing (i.e., $d_i=0$, $\forall i\in{\cal N}$) and offloading (i.e., $\ell_i=0$, $\forall i\in{\cal N}$), respectively. It is observed in Fig.~\ref{fig.vs-d} that the proposed optimal offline designs achieve significant performance gains over the benchmark schemes. This implies the energy-saving benefit of enabling joint task allocation for both local computing and offloading simultaneously. As $d$ increases, it is observed that the energy consumption of the proposed and the full-offloading schemes first increases and then decreases. This is expected, since a large value of $d$ leads to a small channel power gain for WPT but a large channel power gain for offloading. By contrast, the local-computing-only scheme is observed to lead to an increasing energy consumption as $d$ increases, due to the decreasing channel power gain for WPT (as no offloading is employed in this scheme). In addition, the full-offloading scheme is observed to outperform the local-computing-only scheme at large $d$ values (e.g., $d>4$ m in Fig.~\ref{fig.vs-d}(a) and $d>5$ m in Fig.~\ref{fig.vs-d}(b)), but the opposite is true at small $d$ values. Furthermore, by comparing Figs.~\ref{fig.vs-d}(a) and \ref{fig.vs-d}(b), it is observed that in the scenario with time-varying channels, all the three schemes lead to significantly more energy consumption than the corresponding schemes in the scenario with static channels, due to the wireless channel fluctuations over time.

\begin{figure}
  \centering
  \begin{minipage}{8cm}
  \includegraphics[width = 8.0cm]{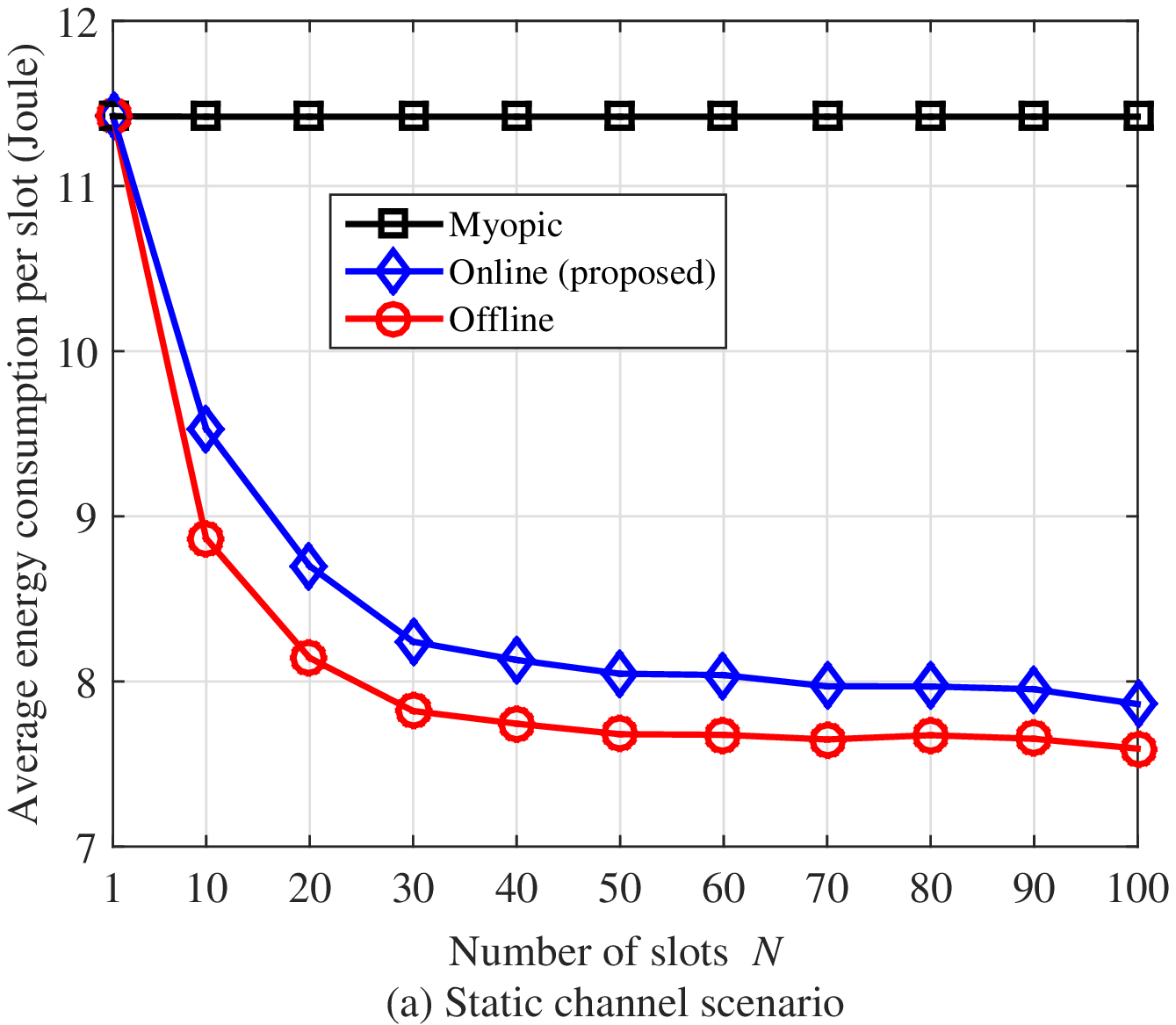}
  \end{minipage}
  \begin{minipage}{8cm}
  \includegraphics[width = 8.0cm]{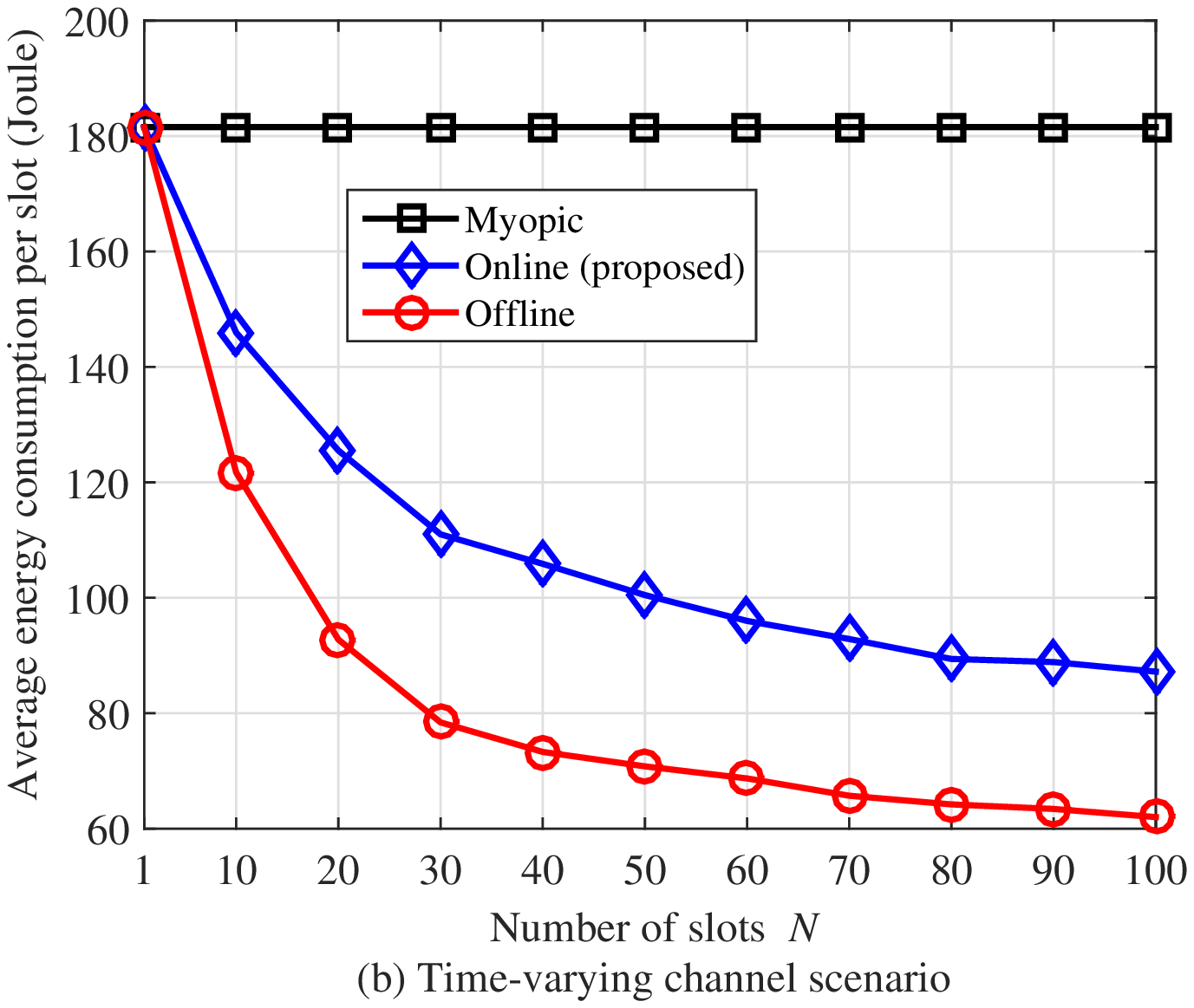}
  \end{minipage}
 \caption{Average energy consumption per slot at the ET versus the number $N$ of slots within the horizon.} \label{fig.vsN-new}
\end{figure}
Next, we consider the online designs in the case with causal CSI/TSI available. For comparison, we consider a benchmark online scheme, namely the myopic design, in which the user needs to accomplish the execution of the arrived tasks at each slot $i\in{\cal N}$. Figs.~\ref{fig.vsN-new}(a) and \ref{fig.vsN-new}(b) show the average transmission energy at the ET per slot versus the number of slots $N$, where $A_{\max}=5\times 10^5$ bits and $d=3$ m. It is observed that the average energy consumption values achieved by the proposed offline and online designs both decrease as $N$ increases, but that by the myopic scheme remains unchanged. This is because our proposed designs can optimize the joint energy and task allocation over time to exploit the time-dynamics in channel fluctuations and task arrivals for energy saving, but the myopic design cannot exploit such time-dynamics. It is also observed that the performance gain achieved by the proposed online designs over the myopic design becomes more significant as $N$ increases, and the proposed designs perform close to the optimal offline designs in both scenarios with static and time-varying channels.

\begin{figure}
  \centering
  \begin{minipage}{8cm}
  \includegraphics[width = 8.0cm]{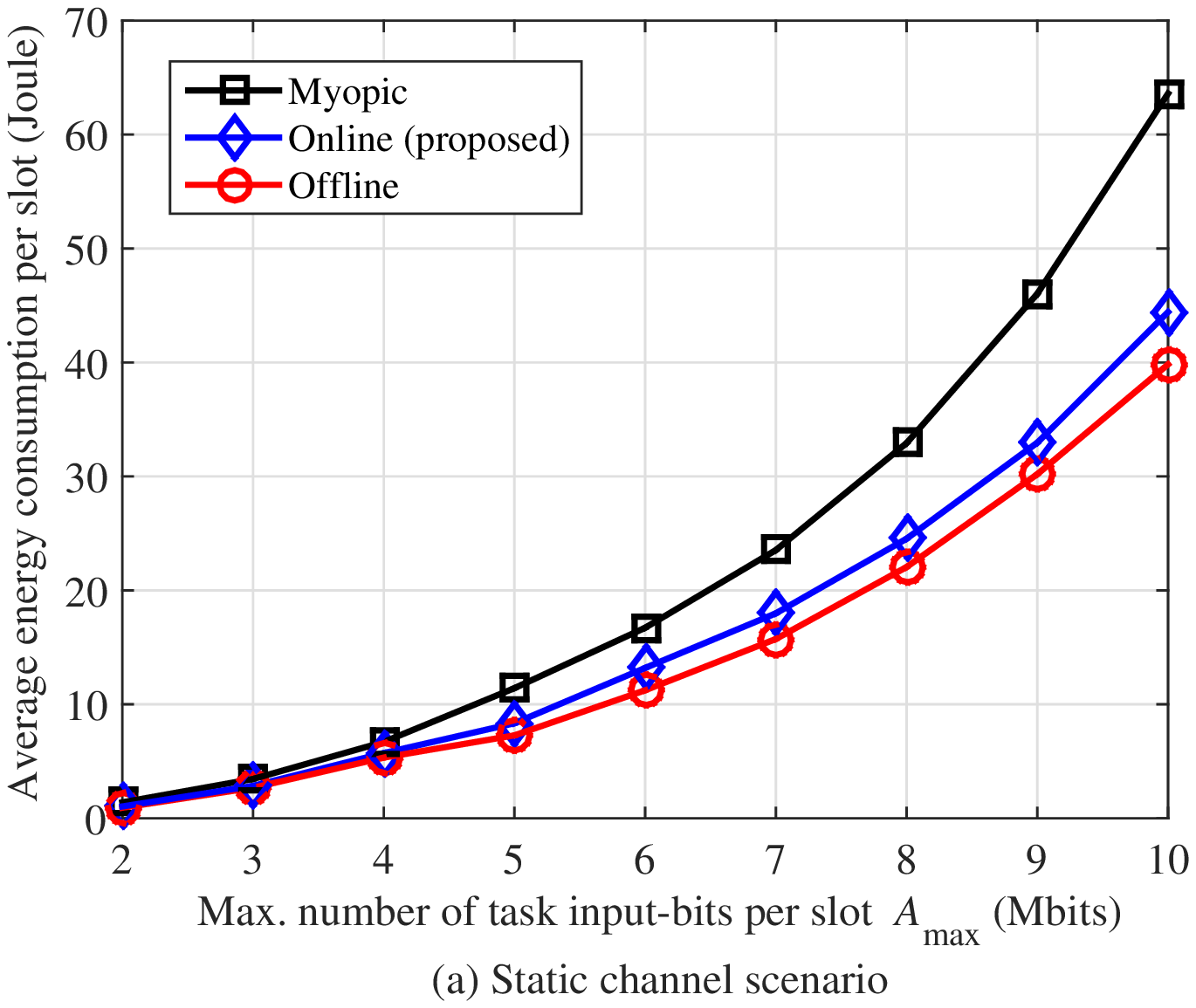}
  \end{minipage}
  \begin{minipage}{8cm}
  \includegraphics[width = 8.0cm]{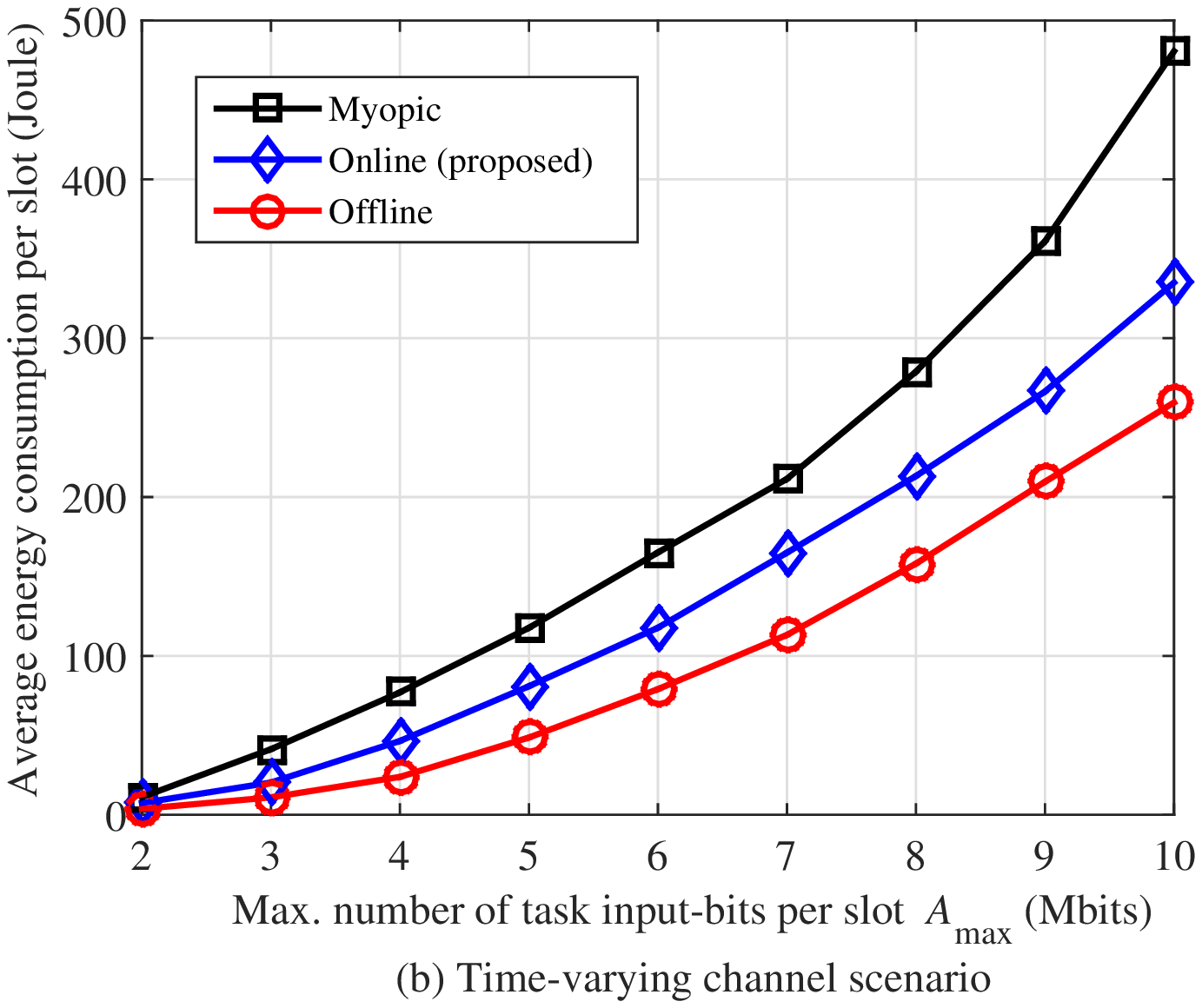}
  \end{minipage}
 \caption{Average energy consumption per slot at the ET versus the maximum number of task input-bits $A_{\max}$.} \label{fig.vsAmax-new}
\end{figure}

Figs.~\ref{fig.vsAmax-new}(a) and \ref{fig.vsAmax-new}(b) show the average transmission energy at the ET per slot versus the maximum number of task input-bits $A_{\max}$ under the scenarios with static and time-varying channels, respectively, where $N=50$ and $d=3$ m. It is observed that the average energy consumption achieved by all the schemes increases as $A_{\max}$ increases. Similarly as Fig.~\ref{fig.vsN-new}, the proposed online designs perform close to the offline designs, and significantly outperform the benchmark myopic designs. The performance gain of the proposed online designs is observed to become more substantial when $A_{\max}$ becomes large. This can be explained similarly as for~Fig.~\ref{fig.vsN-new}.

\section{Concluding Remark}
This paper studied the optimal joint energy and task allocation problem for a single-user wireless powered MEC system with dynamic task arrivals over time, in which we minimize the transmission energy consumption at the ET subject to the energy/task causality and task completion constraints at the user within a finite horizon of multiple slots. Leveraging the convex optimization techniques, we obtained the well-structured optimal offline solutions with non-causal CSI/TSI known {\em a-priori}, in the scenarios with static and time-varying channels, respectively. Inspired by the obtained offline solutions, we further proposed heuristic online joint energy and task allocation designs with only causal CSI/TSI available. Numerical results were provided to show that under both scenarios with static and time-varying channels, the proposed designs achieve significantly smaller energy consumption than benchmark schemes with only local computing or full offloading at the user, and the proposed heuristic online designs perform close to the optimal offline solutions and outperform the conventional myopic designs.

Building on this work, several interesting extensions are worthy of further investigation in the future work for energy-efficient wireless powered MEC designs by considering the nonlinear RF-to-DC conversion, the non-partitionable tasks with binary offloading, the case with individual task computation latency, and the multiple users setting. We briefly discuss these extensions in order as follows.
\begin{itemize}
\item First, it is worth emphasizing that although the linear energy harvesting model is considered, the unified WPT-MEC design principles in this paper are extendible to more general nonlinear energy harvesting models, by replacing the harvested energy function in problem (${\cal P}1$) of interest. For instance, if the harvested DC power is modelled as a sigmoid function with respect to the input RF power (see, e.g., \cite{Clerckx}), then the power allocation problem in problem (${\cal P}1$) will become a non-convex optimization problem that is more challenging to solve optimally. On the other hand, if the signal waveform design for WPT is taken into account (see, e.g., \cite{Zeng17,Clerckx}), then a joint optimization of transmit power allocation and waveform design at the ET should be considered in problem (${\cal P}1$), together with the user's task allocation over time. This will thus make the unified WPT-MEC design problem more complicated.
\item This paper considers the partial offloading case such that each task can be arbitrarily partitioned into two parts for local computing and offloading, respectively. In practice, the binary offloading \cite{Bi18,JunZhang17} is another interesting case that is worth investigation in the future work. In this case, due to the combinatorial nature of the task allocation over time, the formulated energy minimization problem becomes a mixed-integer optimization problem, which is generally NP-hard. As such, other optimization methods such as brunch-and-bound may be applicable to obtain the optimal solution, and deep reinforcement learning may be a feasible method to obtain a high-quality solution with low complexity \cite{Suzhi18}. Alternatively, we can also reuse our proposed solution in the partial offloading case to obtain an approximate solution, by first relaxing the binary offloading variables into continuous ones, obtaining the partial offloading solution, and finally rounding the solution into binary ones. How to choose different methods to solve the unified WPT-MEC problem in different application scenarios should consider the tradeoff between performance and complexity.
\item It is also worth emphasizing that our results are generally extendable to the case with individual task latency, by replacing the single common deadline constraint in (\ref{eq.prob0}d) as several individual latency constraints for different tasks. In this case, since these individual latency constraints do not change the convexity of the energy minimization problem of interest, the similar methods used for solving problem (${\cal P}1$) are still applicable for solving the new problem. Nevertheless, the optimal solution structure (e.g., the monotonically increasing task allocation property) may not hold any more. How to obtain well-structured solution in this case becomes a challenging task that is left for future work. Instead of finding the optimal solution, one potential solution is to simplify this problem by dividing the horizon into several sub-horizons according to the tasks' computation latencies, and then apply the solution in the single-common-deadline case to obtain a high-quality (though sub-optimal in general) solution.
\item Furthermore, although this paper focuses on the case with one single user, the results obtained herein provide useful design guidelines for the more general scenario with multiple users. In the multiuser scenario, the ET needs to optimize the energy beamforming to charge multiple users simultaneously (see, e.g., \cite{Feng18}), and each user needs to adaptively control its task allocation based on the harvested energy from the ET. In this scenario, the task deadline-constrained energy minimization problem can still be formulated as a convex optimization problem similarly as problem (${\cal P}1$), which may be solved by similar methods as adopted in this paper. It is expected that, in order to minimize the energy consumption, the users may still prefer to follow a monotonically increasing energy allocation over time, but the ET may need to design the energy beamforming more intelligently for balancing the energy demands at multiple users.
\end{itemize}

\appendix

\subsection{Proof of Theorem \ref{lem.power_allocation}}\label{proof_lem.power_allocation}
First, we prove that $p_i=0$, $\forall i\in{\cal N}\setminus {\cal N}_{\rm CDS}$ by contradiction. For any energy allocation solution of $\{p_i\}_{i=1}^N$ that satisfies the energy causality constraints in (\ref{eq.prob0}b), we assume that there exists a slot $j\in{\cal N}\setminus{\cal N}_{\rm CDS}$ with $p_j>0$. We then have a CDS $k\in{\cal N}_{\rm CDS}$ such that $1\leq k < j$ and $h_k>h_j$. As such, we can construct another energy allocation solution of $\{\tilde{p}_i\}_{i=1}^N$ by setting $\tilde{p}_k=p_k+\frac{h_j}{h_k}p_j$, $\tilde{p}_j=0$, and $\tilde{p}_m=p_m$, $\forall m\in{\cal N}\setminus\{k,j\}$. It can be verified that the energy allocation of $\{\tilde{p}_i\}_{i=1}^N$ satisfies the energy causality constraints in (\ref{eq.prob0}b). Since $h_k>h_j$ and $p_j>0$, the value $\sum_{i=1}^{N}\tau {\tilde p}_i$ is smaller than $\sum_{i=1}^{N}\tau p_i$. In other words, the energy allocation of $\{{\tilde p}_i\}_{i=1}^N$ achieves a smaller objective value for problem (${\cal P}1$) than $\{p_i\}$. This implies that the energy allocation $\{p_i\}$ is not optimal to problem (${\cal P}1$). Therefore, the optimal energy allocation solution to problem (${\cal P}1$) must satisfy that $p_i=0$, $\forall i\in{\cal N}\setminus {\cal N}_{\rm CDS}$.

Next, we prove that $p_{\phi_k}=\frac{1}{\tau\eta h_{\phi_k}}\sum_{j=\phi_k}^{\phi_{k+1}-1}(E^{\rm loc}(\ell_j)+E_j^{\rm offl}(d_j))$ holds for any CDS $\phi_k$. First, we consider the last CDS $\phi_{|{\cal N}_{\rm CDS}|}$. In this case, by contradiction, we assume that $\tau\eta h_{\phi_{|{\cal N}_{\rm CDS}|}} p_{\phi_{|{\cal N}_{\rm CDS}|}} > \sum_{j=\phi_{|{\cal N}_{\rm CDS}|}}^N(E^{\rm loc}(\ell_j)+E_j^{\rm offl}(d_j))$. Since $\sum_{i=1}^N \tau p_i=\sum_{k=1}^{|{\cal N}_{\rm CDS}|} \tau p_{\phi_k}$, in order to achieve a smaller objective value for problem (${\cal P}1$), we can always decrease the value $p_{\phi_{|{\cal N}_{\rm CDS}|}}$ to ensure that $\tau\eta h_{\phi_{|{\cal N}_{\rm CDS}|}} p_{\phi_{|{\cal N}_{\rm CDS}|}} = \sum_{j=\phi_{|{\cal N}_{\rm CDS}|}}^N(E^{\rm loc}(\ell_j)+E_j^{\rm offl}(d_j))$ holds. Therefore, it follows that $p_{\phi_k}=\frac{1}{\tau\eta h_{\phi_k}}\sum_{j=\phi_k}^{\phi_{k+1}-1}(E^{\rm loc}(\ell_j)+E_j^{\rm offl}(d_j))$ at CDS $\phi_{|{\cal N}_{\rm CDS}|}$. Then, for any CDS $\phi_k$, $k\in \{1,\ldots,|{\cal N}_{\rm CDS}|-1\}$, we assume that the energy allocation $p_{\phi_k}$ does not satisfy $p_{\phi_k}=\frac{1}{\tau\eta h_{\phi_k}}\sum_{j=\phi_k}^{\phi_{k+1}-1}(E^{\rm loc}(\ell_j)+E_j^{\rm offl}(d_j))$. It then follows that $\Delta_{\phi_k}>0$, where $\Delta_{\phi_k}\triangleq \tau\eta h_{\phi_k} p_{\phi_k} - \sum_{j=\phi_k}^{\phi_{k+1}-1}( E^{\rm loc}(\ell_j)+E_j^{\rm offl}(d_j))$ is defined. We again construct one feasible energy allocation $\{\tilde{p}_i\}_{i=1}^N$ for problem (${\cal P}1$) by setting $\tilde{p}_{\phi_k}= p_{\phi_k} - \frac{\Delta_{\phi_k}}{\tau \eta h_{\phi_k}}$, $\tilde{p}_{\phi_{k+1}}= p_{\phi_{k+1}} + \frac{\Delta_{\phi_k}}{\tau \eta h_{\phi_{k+1}}}$, and $\tilde{p}_m=p_m$, $\forall m\in{\cal N}\setminus\{\phi_{k},\phi_{k+1}\}$. It is verified that the new energy allocation of $\{\tilde{p}_i\}_{i=1}^N$ satisfies the constraints in (\ref{eq.prob0}b). Since $\Delta_{\phi_k}>0$ and $h_{\phi_{k+1}}>h_{\phi_{k}}$, we have $\sum_{i=1}^N \tau \tilde{p}_i<\sum_{i=1}^N \tau p_i$, which implies that the energy allocation of $\{\tilde{p}_i\}_{i=1}^N$ achieves a smaller objective value than $\{p_i\}_{i=1}^N$ for problem (${\cal P}1$). This contradicts the assumption that $\{p_i\}$ is optimal for problem (${\cal P}1$). Therefore, it must hold that $\Delta_{\phi_k}=0$, $\forall k\in\{1,\ldots,|{\cal N}_{\rm CDS}|-1\}$, and we now complete the proof of $p_{\phi_k}=\frac{1}{\tau\eta h_{\phi_k}}\sum_{j=\phi_k}^{\phi_{k+1}-1}(E^{\rm loc}(\ell_j)+E_j^{\rm offl}(d_j))$ for all CDSs $\phi_k$'s with $k\in\{1,\ldots,|{\cal N}_{\rm CDS}|\}$. As a result, Theorem 2 is finally verified.

\end{document}